\documentclass[11pt,a4paper]{article}
\pdfoutput=1

\usepackage{a4wide}
\usepackage{graphicx} 
\usepackage{amsmath}
\usepackage{amssymb}
\usepackage{amsfonts}
\usepackage{cite}
\usepackage{color}
\usepackage{adjustbox}
\usepackage{pdflscape}
\usepackage{gensymb}
\usepackage{graphicx}
\usepackage{diagbox}
\usepackage{pifont}
\usepackage{bigstrut}
\usepackage[small,bf]{caption}
\usepackage{enumerate}
\usepackage{tabulary}
\usepackage{appendix}
\usepackage{hyperref}
\usepackage{makecell}
\usepackage{multirow}
\usepackage{dcolumn}
\usepackage{textcomp}

\usepackage{ae} 
\usepackage{slashed}
\usepackage{aecompl}
\usepackage{bm}

\usepackage{epsfig}
\usepackage{fontenc}
\usepackage{graphicx}
\usepackage[lofdepth,lotdepth,caption=false]{subfig}
\usepackage{color}
\usepackage{booktabs}
\usepackage{tabularx}
\usepackage{multirow}
\usepackage[table]{xcolor}
\usepackage[normalem]{ulem}
\usepackage{dcolumn}
\usepackage{rotating}
\usepackage{hyperref}
\usepackage{multirow,bigdelim}  

\usepackage[latin1]{inputenc} 

\newcommand{\tzm}{\ensuremath{\theta_{23}}}

\newcommand{\ie}{\textit{i.e.}}

\newcommand{\dcp}{\delta_{\mathrm{CP}}}
\newcommand{\Amue}{\mathcal{A}^{\mu e}_{\mathrm{CP}}}
\newcommand{\Amuevac}{[\mathcal{A}^{\mu e}_{\mathrm{CP}}]_{\mathrm{vac}}}
\newcommand{\Amuemat}{[\mathcal{A}^{\mu e}_{\mathrm{CP}}]_{\mathrm{mat}}}

\newcommand{\capdef}{}
\newcommand{\mycaption}[2][\capdef]{\renewcommand{\capdef}{#2}
	\caption[#1]{{\footnotesize #2}}}

\begin{document}

\begin{titlepage}

\vspace*{-15mm}

\begin{flushright}
IP/BBSR/2022-08
\end{flushright}

\vspace*{0.8cm}

\begin{center}

{\bf\Large{Enhancing Sensitivity to Leptonic CP Violation using Complementarity among DUNE, T2HK, and T2HKK}} \\ [10mm]

{\bf Sanjib Kumar Agarwalla,}$^{\, a,b,c,d,}$\footnote{E-mail: \texttt{sanjib@iopb.res.in, (ORCID:0000-0002-9714-8866)}}, 
{\bf Sudipta Das,}$^{\, a,b,}$\footnote{E-mail: \texttt{sudipta.d@iopb.res.in, (ORCID:0000-0002-5508-7751)}}, 
{\bf Alessio Giarnetti,}$^{\, e,}$\footnote{E-mail: \texttt{giarnetti.alessio@unioma3.it, (ORCID:0000-0001-8487-8045)}},\\ 
{\bf Davide Meloni,}$^{\, e,}$\footnote{E-mail: \texttt{davide.meloni@uniroma3.it, (ORCID:0000-0001-7680-6957)}}
{\bf Masoom Singh}$^{\, a,f,}$\footnote{E-mail: \texttt{masoom@iopb.res.in, (ORCID:0000-0002-8363-7693)}} \\
\vspace{8mm}
$^{a}$\,{\it Institute of Physics, Sachivalaya Marg, Sainik School Post, Bhubaneswar 751005, India}\\
\vspace{2mm}
$^{b}$\,{\it Homi Bhabha National Institute, Training School Complex, Anushakti Nagar, Mumbai 400094, India}\\
\vspace{2mm}
$^{c}$\,{\it International Centre for Theoretical Physics, Strada Costiera 11, 34151 Trieste, Italy}\\
\vspace{2mm}
$^{d}$\,{\it Department of Physics \& Wisconsin IceCube Particle Astrophysics Center, University of Wisconsin, Madison, WI 53706, U.S.A}\\
$^{e}$\,{\it Dipartimento di Matematica e Fisica, Universit\`a di Roma Tre Via della Vasca Navale 84, 00146 Rome, Italy}\\	
$^{f}$\,{\it Department of Physics, Utkal University, Vani Vihar, Bhubaneswar 751004, India}

\end{center}

\vspace{7mm}

\begin{abstract}
\vspace{4mm}
\noindent 
After the landmark discovery of non-zero $\theta_{13}$ by the modern reactor experiments,
unprecedented precision on neutrino mass-mixing parameters has been achieved over the past decade.
This has set the stage for the discovery of leptonic CP violation (LCPV) at high confidence level in the
next-generation long-baseline neutrino oscillation experiments. In this work, we explore in detail the
possible complementarity among the on-axis DUNE and off-axis T2HK experiments to enhance the
sensitivity to LCPV suppressing the $\theta_{23}-\delta_{\mathrm{CP}}$ degeneracy. We find that none of these
experiments individually can achieve the milestone of 3$\sigma$ LCPV for at least 75\% choices
of $\delta_{\mathrm{CP}}$ in its entire range of $[-180^{\circ} , 180^{\circ}]$, with their nominal exposures and
systematic uncertainties. However, their combination can attain the same for all values of $\theta_{23}$
with only half of their nominal exposures. We observe that the proposed T2HKK setup in combination
with DUNE can further increase the CP coverage to more than 80\% with only half of their nominal
exposures. We study in detail how the coverage in $\delta_{\mathrm{CP}}$ for $\ge$ 3$\sigma$ LCPV depends
on the choice of $\theta_{23}$, exposure, optimal runtime in neutrino and antineutrino modes,
and systematic uncertainties in these experiments in isolation and combination. We find that
with an improved systematic uncertainty of 2.7\% in appearance mode, the standalone T2HK
setup can provide a CP coverage of around 75\% for all values of $\theta_{23}$. We also discuss
the pivotal role of intrinsic, extrinsic, and total CP asymmetries in the appearance channel and
extrinsic CP asymmetries in the disappearance channel while analyzing our results.
\end{abstract}

\end{titlepage}

\setcounter{footnote}{0}




\section{Introduction and Motivation}
\label{sec:1}

One of the fundamental properties of particles is their behavior under the 
CP (charge-parity) transformation and a violation of the CP symmetry may
have an important connection to the observed baryon asymmetry 
in the Universe~\cite{Sakharov:1967dj}. So far, in the quark 
sector of the Standard Model (SM), we have two known sources of  
CP-invariance violation~\cite{ParticleDataGroup:2022pth}. One of them is the 
CP-odd phase in the Cabibbo-Kobayashi-Maskawa (CKM) matrix,
which is known to be large and governs all the CP-violating phenomena
observed so far. The other one is the so-called strong CP-phase 
$\theta_{\rm QCD}$, which is known to be vanishingly small. In the 
lepton sector, we achieved an important breakthrough in 2012
in establishing the standard three-flavor oscillation picture of neutrinos
through the pioneering discovery of the non-zero value of the smallest neutral lepton mixing
angle $\theta_{13}$ by the Daya Bay reactor antineutrino 
experiment~\cite{DayaBay:2012fng}. This landmark finding 
opened the door for a completely new and independent source of 
CP invariance violation in neutrino oscillation experiments. 
The so-called Dirac CP-odd phase $\dcp$ in the $3 \times 3$ 
unitary Pontecorvo-Maki-Nakagawa-Sakata (PMNS) matrix 
is the source of CP-invariance violation in the neutral lepton sector,
which can be probed via neutrino oscillation probabilities. 

After the discovery of non-zero $\theta_{13}$, remarkable precision has 
been achieved on neutrino mass-mixing parameters over the past decade,
which has enabled us to come up with a simple, robust, three-flavor neutrino
oscillation paradigm, which is capable of accommodating most of the oscillation
data~\cite{deSalas:2020pgw,Esteban:2020cvm,NuFIT,Capozzi:2021fjo}.
In 3$\nu$ oscillation picture, if the value of $\dcp$ turns out to be different 
from both $0^{\circ}$ and $180^{\circ}$ in Nature, then it would cause 
a difference between neutrino and antineutrino transition probabilities -- 
providing a smoking gun signature of CP violation (CPV) in neutrino 
oscillation experiments. In the intensity frontier, the currently running
and upcoming high-precision long-baseline (LBL) neutrino oscillation 
experiments are the most promising avenues to unravel the novel
signatures of CPV. One of the prime scientific goals of these LBL 
experiments is to provide an explicit demonstration of leptonic CPV 
by precisely measuring the differences between the $\theta_{13}$-driven 
oscillations of muon-type neutrinos and antineutrinos into electron-type 
neutrinos and antineutrinos, respectively.

By observing these differences, the currently running LBL experiments 
T2K~\cite{T2K:2019bcf,T2K:2021xwb} and NOvA~\cite{NOvA:2021nfi} 
have already started probing the parameter space of $\dcp$ and 
provided hints toward non-zero CPV. Now, it bestows upon the 
next-generation LBL experiments in the neutrino roadmap to convert 
these crucial hints of leptonic CPV into discoveries at high confidence 
level. Such a path-breaking discovery would certainly pave the way 
to elucidate the age-old flavor puzzle and the prevalence of matter 
over antimatter in the Universe~\cite{Sakharov:1967dj,Pascoli:2006ie,Branco:2011zb,Petcov:2014laa,Hagedorn:2017wjy}.

In this paper, after having an insightful discussion on the critical role of intrinsic, 
extrinsic and total CP asymmetries in the appearance channel and extrinsic 
CP asymmetries in the disappearance channel, we study in detail the capabilities 
of the next-generation long-baseline neutrino oscillation experiments DUNE
(Deep Underground Neutrino Experiment)~\cite{DUNE:2015lol,DUNE:2020lwj,DUNE:2020ypp,DUNE:2020jqi,DUNE:2021cuw,DUNE:2021mtg}
and T2HK (Tokai to Hyper-Kamiokande)~\cite{Hyper-KamiokandeProto-:2015xww,Hyper-Kamiokande:2018ofw} in isolation and combination to establish the leptonic CPV ($\dcp$ $\neq$ $0^{\circ}$ and $180^{\circ}$) 
at 3$\sigma$ confidence level for at least 75\% choices\footnote{We often mention this performance 
indicator of a given experiment as ``CP coverage", which denotes the values of true $\dcp$ (in \%) in its entire range of $[-180^{\circ} , 180^{\circ}]$,
for which leptonic CPV can be established at $\ge$ 3$\sigma$ confidence level.}
of true $\dcp$ in its entire range of $-180^{\circ}$ to $180^{\circ}$, considering the 
state-of-the-art simulation details of these facilities. We extend our analysis to the 
proposed T2HKK setup~\cite{Hyper-Kamiokande:2016srs} 
and explore its CP coverage in standalone mode and also in combination with DUNE.
The main thrust of this paper is to investigate in detail the possible complementarity 
among these high-precision experiments~\cite{Fukasawa:2016yue,Ballett:2016daj,Liao:2016orc,Ghosh:2017ged,Choubey:2017cba,Agarwalla:2021owd}
to fully exploit the three-flavor interference effects~\cite{Agarwalla:2013hma} suppressing the parameter degeneracies, 
which in turn enhances the CP coverage. The key point of our analysis is to 
demonstrate how these experiments bring complementary information 
on the CP phase $\dcp$ for different octant choices of the 2-3 mixing angle 
($\theta_{23}$) at different $L/E$ values by means of appearance and disappearance 
channels in neutrino and antineutrino modes. Such a combination is also crucial 
to tackle the underlying degeneracies among $\theta_{23}$ and $\dcp$, 
which reduce the CP coverage in $\dcp$~\cite{Nath:2015kjg,Machado:2015vwa} 
while establishing the leptonic CPV.

The upcoming DUNE is planning to use a 40 kt liquid argon time projection chamber
(LArTPC) as a far detector which will be exposed to an on-axis, high-intensity, wide-band 
neutrino beam covering both the first and second oscillation maxima with a baseline of 1300 km.
The DUNE far detector is expected to have an unmatched kinematic reconstruction capability 
for all the observed particles in the final state, which plays an important role to reject a large 
fraction of the neutral current background. The presence of an efficient near detector will 
significantly minimize the impact of flux and cross-section related systematic uncertainties
at the DUNE far detector. DUNE will experience a significant amount of Earth's matter effect
because of its large baseline of 1300 km and having access to a wide-band
beam whose flux extends up to $\sim$ 6 GeV with a peak at around 2.5 GeV. All these features 
help DUNE to settle the issue of
neutrino mass ordering\footnote{The sign of $\Delta m^{2}_{31}$ ($\equiv m_3^2 - m_1^2$) 
is still unknown. If $\Delta m^{2}_{31} > 0$, it is known as normal mass ordering (NMO) 
($m_{3} \gg m_{2} > m_{1}$), while if $\Delta m^{2}_{31} < 0$, it is referred to as inverted 
mass ordering (IMO) ($m_{2} > m_{1} \gg m_{3}$).} at a very high confidence level irrespective of the choices of other 
oscillation parameters and to measure the values of $\dcp$ and $\theta_{23}$ with satisfactory 
precision utilizing the information on oscillation pattern at several $L/E$ values~\cite{Arafune:1997hd,DUNE:2020ypp,DUNE:2020jqi}.
On the other hand, in Japan, the proposed gigantic 187 kt Hyper-Kamiokande (HK) 
water Cherenkov detector will serve as the far detector for the T2HK (JD) setup at a distance of
295 km from the J-PARC facility and receive an off-axis ($2.5^{\circ}$), upgraded, 
narrow-band beam with a flux peaking around the first oscillation maximum of $\sim$ 0.6 GeV.
The shorter baseline and high statistics help T2HK (JD) to achieve an unparalleled
precision in the measurement of $\dcp$ and $\theta_{23}$ free from Earth's matter 
effect~\cite{Hyper-KamiokandeProto-:2015xww,Hyper-Kamiokande:2018ofw}. 
The Korean detector (KD) with a baseline of 1100 km and an off-axis beam from 
J-PARC having a peak around the second oscillation maximum is slightly sensitive 
to Earth's matter effect and provides complementary information on $\dcp$ as 
compared to JD~\cite{Hyper-Kamiokande:2016srs}.

In this paper, we study in detail how the coverage in $\dcp$ for $\ge$ 3$\sigma$ 
leptonic CPV varies with the choice of $\theta_{23}$, exposure, optimal runtime 
in neutrino and antineutrino modes, and systematic uncertainties in these experiments 
in isolation and combination. We find that neither DUNE nor T2HK individually can achieve 
the milestone of 75\% CP coverage for which at least 3$\sigma$ leptonic CPV
can be ensured with their nominal exposures and systematic uncertainties.
In DUNE, we observe that the main bottleneck is $\theta_{23}-\dcp$ 
degeneracy which appears in the picture when $\theta_{23}$ lies in the 
range of 42$^{\circ}$ to 48$^{\circ}$. We notice that this degeneracy cannot
be resolved in DUNE even by doubling the exposure or reducing the current 
systematic uncertainties by a factor of two. While in T2HK, although such 
degeneracy does not play any significant role because of the negligible 
matter effects, the current systematic uncertainties are an obstacle in achieving 
the above-mentioned sensitivity. One of the important conclusions of our work 
is that the complementarity between DUNE and T2HK is essential to obtain the
desired CP coverage irrespective of the value of $\theta_{23}$ in Nature.
Our study shows that for the combination of DUNE and T2HK, only half of their 
nominal exposures are sufficient to establish 3$\sigma$ leptonic CPV for at least 
75\% choices of $\dcp$ for almost all values of $\theta_{23}$, with their nominal 
systematic uncertainties. This becomes possible due to the less systematic 
uncertainties in DUNE as compared to T2HK and high matter-independent 
disappearance statistics in T2HK, that helps in constraining $\theta_{23}$ 
in a narrow range and thus to resolve the $\theta_{23}-\dcp$ degeneracy.
We find that with an improved systematic uncertainty of 2.7\% in appearance 
mode, the standalone T2HK (JD) setup can provide a CP coverage of around 75\% 
for almost all values of $\theta_{23}$ with nominal exposure. We observe that 
with nominal exposure and systematic uncertainties, T2HKK (JD+KD) 
can also achieve the 75\% CP coverage for all values of $\theta_{23}$, 
but its CP coverage is always less than that of DUNE+JD.
At the same time, with only half of their exposures and nominal systematic 
uncertainties, T2HKK+DUNE can achieve a CP coverage of more than 
80\% for almost all values of $\theta_{23}$.

We organize the paper as follows. We initiate our discussion with a detailed analytical 
understanding of CP asymmetry in Sec.~\ref{sec:2}, which will come in handy while 
analyzing our findings. In Sec.~\ref{sec:3}, we outline our experimental and simulation 
details of different setups under consideration. Sec.~\ref{sec:4}  summarizes our results 
and findings wherein we discuss our milestone of achieving 3$\sigma$ leptonic CPV
for at least 75\% choices of $\dcp$ in these experiments in isolation and combination 
as a function of (a) true $\sin^{2}\theta_{23}$ in Sec.~\ref{sec:4a}, 
(b) varying exposure in Sec.~\ref{sec:4b}, 
(c) optimal runtime in neutrino and antineutrino modes in Sec.~\ref{sec:4c}, 
and (d) systematic uncertainties in Sec.~\ref{sec:4d}.
Also, in Sec.~\ref{sec:4e}, we estimate the enhanced CP coverage of these
experiments when we assume the values of true $\dcp$ only in its current 
3$\sigma$ allowed range of $[-175^{\circ} , 41^{\circ}]$, instead of its entire
range of $[-180^{\circ} , 180^{\circ}]$. In Sec.~\ref{sec:5}, we discuss the results we would expect if the neutrino mass ordering is inverted.
Finally, we summarize with our concluding remarks in Sec.~\ref{sec:6}.

\section{Discussion at the Oscillation Probability Level }
\label{sec:2}
%
The mixing matrix in the standard three-neutrino (3$\nu$) framework is written in terms of the three mixing angles ($\theta_{23}$, $\theta_{13}$, and $\theta_{12}$) 
and one complex phase ($\delta_{\mathrm{CP}}$)~\cite{Zyla:2020zbs}. Following the usual PMNS parameterization, we have :
\begin{eqnarray}
   \nonumber  U_{\mathrm{PMNS}}&=
     \begin{pmatrix}
    c_{12}c_{13} & s_{12}c_{13} & s_{13}e^{-i\delta_{\mathrm{CP}}} \\
     -s_{12}c_{23}-c_{12}s_{23}s_{13}e^{i\delta_{\mathrm{CP}}} & c_{12}c_{23}-s_{12}s_{23}s_{13}e^{i\delta_{\mathrm{CP}}} & s_{23}c_{13} \\
     s_{12}s_{23}-c_{12}c_{23}s_{13}e^{i\delta_{\mathrm{CP}}} & -c_{12}s_{23}-s_{12}c_{23}s_{13}e^{i\delta_{\mathrm{CP}}} & c_{23}c_{13}
     \end{pmatrix}\,,
 \end{eqnarray}
 where we notice that the Dirac phase $\delta_{\mathrm{CP}}$ is always coupled to the mixing angles. This explains that sensitivity in the CP phase strongly depends on the knowledge of other mixing parameters. 
At LBL experiments, we mostly probe the $\nu_\mu (\bar{\nu}_{\mu})\to\nu_{\mu} (\bar{\nu}_{\mu})$ (disappearance) and the $\nu_{\mu} (\bar{\nu}_{\mu})\to\nu_e (\bar{\nu}_{e})$ (appearance) channels. Following the approach in Ref.~\cite{Agarwalla:2021bzs}, we can further simplify the appearance probability expression in Ref.~\cite{Akhmedov:2004ny} dropping the small $\alpha^2$ terms, where $\alpha=\Delta m_{21}^2/\Delta m_{31}^2$\,, as follows: 
\begin{eqnarray}
  P_{\mu e}\approx N\sin^2\theta_{23}+O\sin2\theta_{23}\cos(\Delta+\delta_{\mathrm{CP}})\,,
  \label{eq:1}
\end{eqnarray}
where,
\begin{eqnarray}
  N&=&4\sin^2\theta_{13}\frac{\sin^2[(\hat A-1)\Delta]}{(\hat A-1)^2}\, ,
  \label{eq:2}
  \end{eqnarray}
  \begin{eqnarray}
  O&=&2\alpha\sin\theta_{13}\sin2\theta_{12}\frac{\sin\hat A\Delta}{\hat A}\frac{\sin[(\hat A-1)\Delta]}{\hat A-1}\,.
  \label{eq:3}
\end{eqnarray}
This grouping of terms helps to visualize the dependence on the atmospheric mixing angle ($\theta_{23}$). In the above set of equations, $\Delta=\Delta m_{31}^2 L/4E$, and $\hat A=A/\Delta m_{31}^2$, wherein the Wolfenstein matter term, $A=2\sqrt{2}G_F N_e E \approx 2\times7.6\times Y_{e} \times 10^{-5} \times \rho_{\mathrm{avg}} \,$ (g/cm$^3$) $\times E $\, (GeV). Here $\rho_{\mathrm{avg}}$ is the line-averaged constant Earth matter density which we consider as 2.848 g/cm$^3$, 2.7 g/cm$^3$, and 2.8 g/cm$^3$ in DUNE~\cite{Roe:2017zdw}, JD~\cite{Hyper-Kamiokande:2018ofw}, and KD~\cite{Hyper-Kamiokande:2016srs}, respectively. Also, assuming that Earth's matter is electrically neutral and isoscalar, we obtain $N_{e} = N_{p} = N_{n}$ where  $N_{p}$ and $N_{n}$ are the proton and neutron number densities in Earth, respectively. Thus, the relative number density given by: $Y_{e} \equiv N_{e}/(N_{p}+ N_{n})$ is estimated as 0.5. Further, from Eq.~\ref{eq:1}, we observe that the CP-violating term contains $\sin2\theta_{23}$ and thus is insensitive to the octant of the atmospheric angle~\cite{Fogli:1996pv, Barger:2001yr, Minakata:2002qi, Minakata:2004pg,Hiraide:2006vh,Das:2017fcz}. Moreover, changing from neutrino to antineutrino mode notably changes signs of $\hat A$, thus leading to matter-induced or fake (extrinsic) CPV~\cite{Tanimoto:1998sn,Barger:1980tf,Arafune:1997hd}. This will have a dominant contribution as it is present in the coefficient of the leading term: N (refer to Eq.~\ref{eq:2}). While the presence of the Dirac CP phase in the sub-leading term gives rise to the genuine (intrinsic) CPV (refer to Eq.~\ref{eq:3}).

\begin{table}[htb!]
\resizebox{\columnwidth}{!}{%
  \centering
  \begin{tabular}{|c|c|c|c|c|c|c|}
    \hline \hline 
    \multirow{2}{*}{$\sin^2 \theta_{12}$} & \multirow{2}{*}{$\sin^2\theta_{23}$} & \multirow{2}{*}{$\sin^2 \theta_{13}$} &
    $\Delta m^2_{31}$ (eV$^2$) & $\Delta m^2_{21}$ (eV$^2$) & $\delta_{\rm CP}$
    & Mass \\
    & & & $\times10^{-3}$ & $\times10^{-5}$ & ($^{\circ}$) & Ordering\\
    \hline \hline
    0.303 & [0.4, 0.6] & 0.0223 & 2.522 (-2.418) & 7.36
    & [- 180, 180] & NMO (IMO) \\
    \hline \hline 
  \end{tabular}}
  \mycaption{The benchmark values of six oscillation parameters used in our analysis assuming normal mass ordering (NMO). While going from NMO to inverted mass ordering (IMO) (in Sec.~\ref{sec:5}), we change the value of $\Delta m^{2}_{31}$ following Ref.~\cite{Capozzi:2021fjo}.}
  \label{table:one}
\end{table}
%

To probe CPV in oscillation experiments, one needs to note the difference between neutrino and antineutrino oscillation probabilities. The quantity which is strongly correlated to the sensitivity in $\delta_{\mathrm{CP}}$ is the CP asymmetry~\cite{Bernabeu:2019npc,Bernabeu:2018use,Ohlsson:2014cha,Nunokawa:2007qh}. In the following subsections, we will discuss how CP asymmetries affect CPV at the probability level. Throughout our simulation, we use the constant values of $\theta_{12}, \theta_{13}, \Delta m^{2}_{21},$ and $\Delta m^{2}_{31}$ as our benchmark values (refer to Table~\ref{table:one}), assuming NMO. We consider IMO only in Sec.~\ref{sec:5} and Appendix~\ref{sec:appendix1}. We often vary $\theta_{23}$ in some cases, as will be mentioned wherever necessary, while $\dcp$ is always varied in its entire range, except for Sec~\ref{sec:4e} where we make use of the current 3$\sigma$ constraints.

\subsection{Extrinsic, Intrinsic, and Total CP Asymmetries in Appearance Channel}
\label{sec:2a}
\begin{figure}[htb!]
    \centering
    \includegraphics[scale = 0.97]{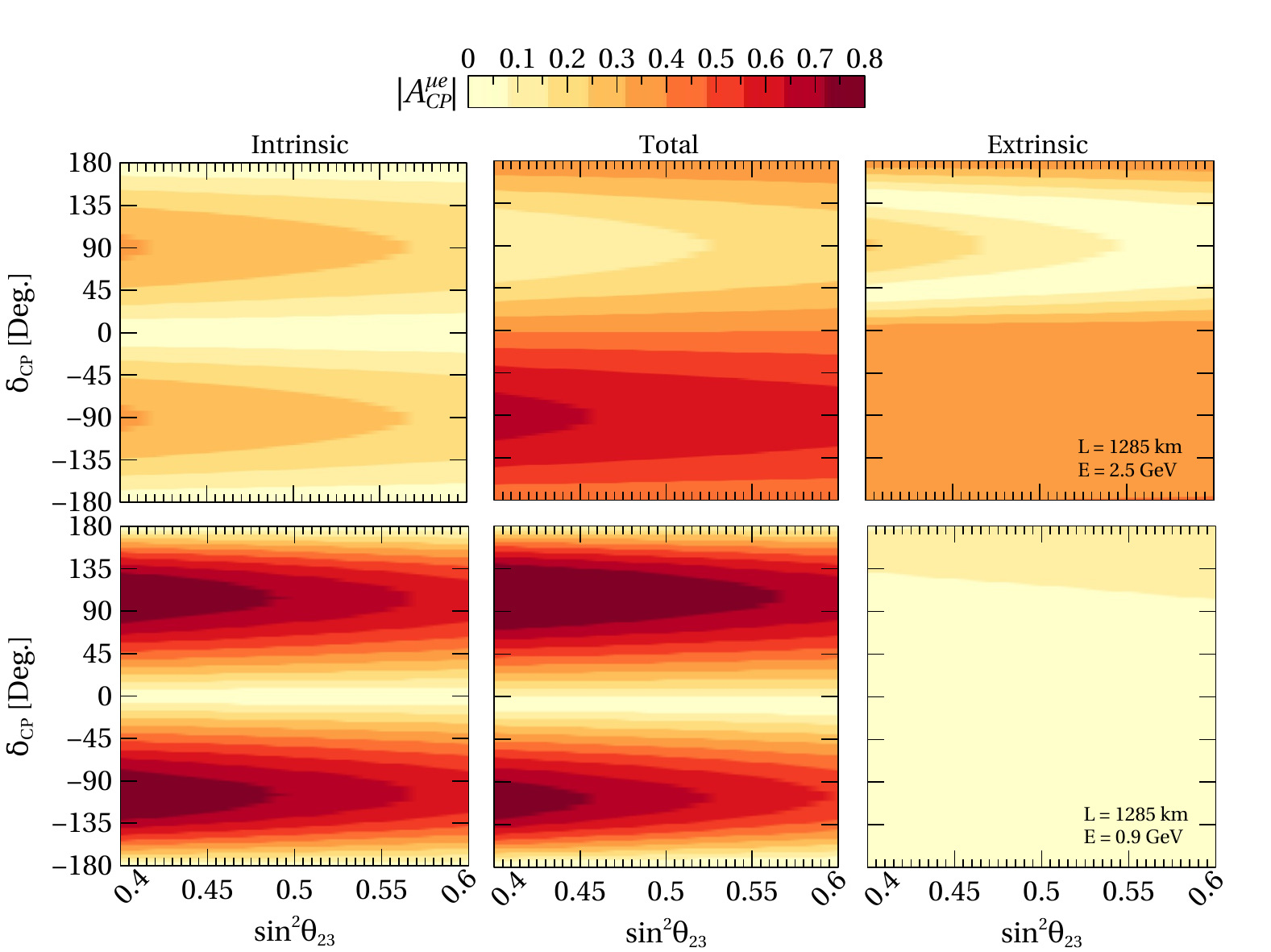}
    \mycaption {Absolute CP asymmetry ($|\Amue|$) as a function of $\delta_{\mathrm{CP}}$ and $\sin^{2}\theta_{23}$ for first oscillation maximum ($L$ = 1285 km, $E$ = 2.5 GeV) and second oscillation maximum ($L$ = 1285 km, $E$ = 0.9 GeV) in DUNE, assuming NMO are shown in the top and bottom panels, respectively. The left and middle panels in both top and bottom are obtained in a vacuum (intrinsic or genuine $\mathcal{A}^{\mu e}_{\mathrm{CP}}$) and finite matter density (both intrinsic and extrinsic $\mathcal{A}^{\mu e}_{\mathrm{CP}}$) scenarios, respectively, while the right panel represents the difference between the first two (only extrinsic or fake $\mathcal{A}^{\mu e}_{\mathrm{CP}}$). Values of other oscillation parameters are taken from Table~\ref{table:one}.}
    \label{fig:1}
\end{figure}
\begin{figure}[htb!]
    \centering
    \includegraphics[scale = 0.97]{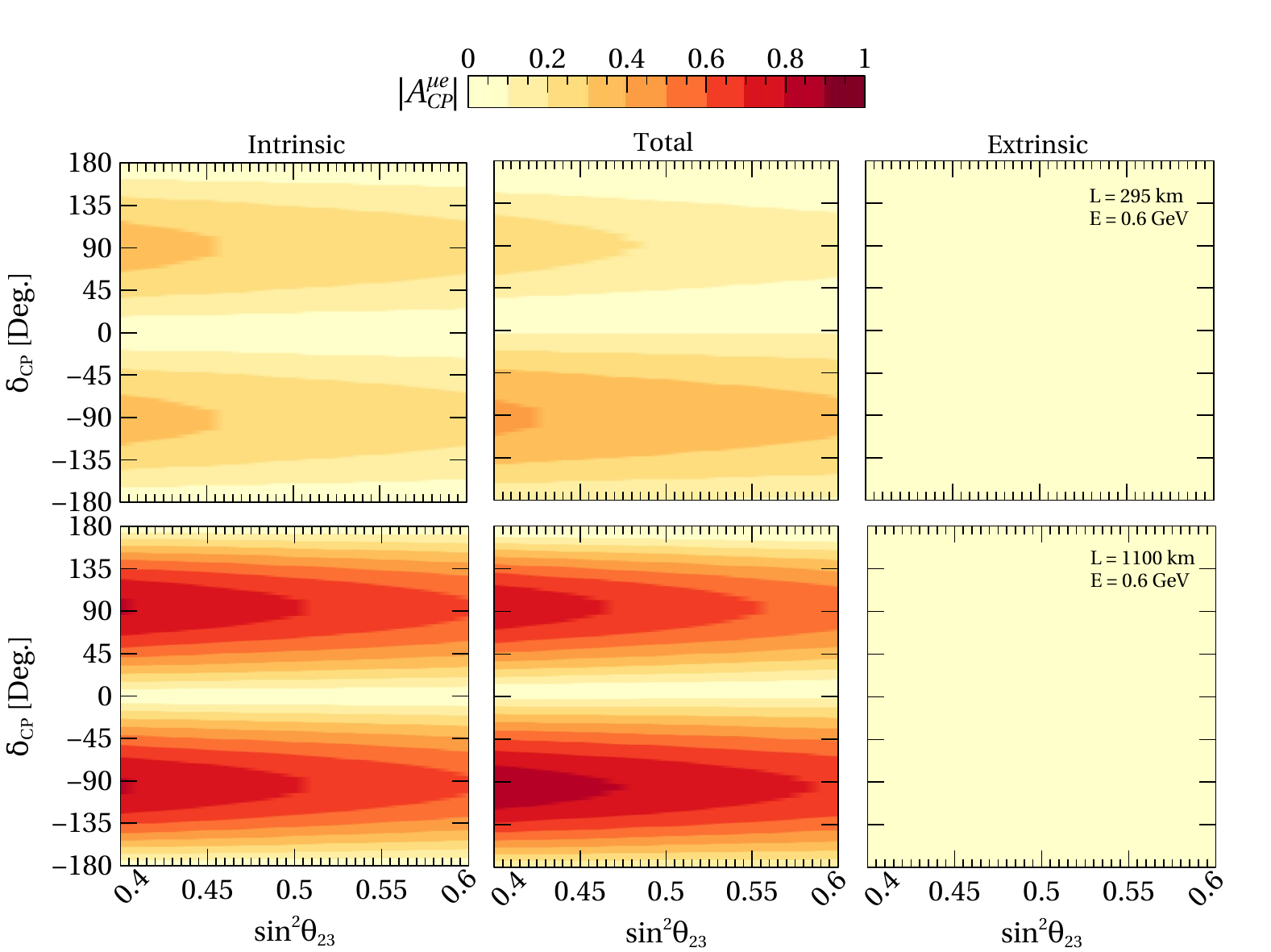}
    \mycaption{$|\mathcal{A}^{\mu e}_{\mathrm{CP}}|$ as a function of $\delta_{\mathrm{CP}}$ and $\sin^{2}\theta_{23}$ for first oscillation maximum in JD ($L$ = 295 km, $E$ = 0.6 GeV) and second oscillation maximum in KD ($L$ = 1100 km, $E$ = 0.6 GeV) assuming NMO are shown in the top and bottom panels, respectively. The left and middle panels in both top and bottom are obtained in vacuum (intrinsic or genuine $\mathcal{A}^{\mu e}_{\mathrm{CP}}$) and finite matter density (intrinsic $\mathcal{A}^{\mu e}_{\mathrm{CP}}$ + extrinsic $\mathcal{A}^{\mu e}_{\mathrm{CP}}$) scenarios, respectively, while the right panel represents the difference between the first two (only extrinsic or fake $\mathcal{A}^{\mu e}_{\mathrm{CP}}$). Values of other oscillation parameters are taken from Table~\ref{table:one}.}
    \label{fig:2}
\end{figure}

The CP asymmetry in the appearance channel is defined as
\begin{eqnarray}
\label{eq:CPA_app}
 \Amue=\dfrac{P_{\mu e}-\bar P_{\mu e}}{P_{\mu e}+\bar P_{\mu e}}\,.
\end{eqnarray}
Different expansions have been done to understand the behavior of such asymmetry in terms of mixing angles \cite{Giarnetti:2021wur}. However, to realize the role of the atmospheric mixing angle in the $\delta_{\mathrm{CP}}$ sensitivity, we fix the remaining mixing angles at their benchmark values ($\sin\theta_{13}\sim 1/7$ and $\sin\theta_{12}\sim 1/\sqrt{3}$). Also, since the value of matter parameter in the considered LBL experiments is not large, we can expand in $\hat A$ up to the first order. The resulting asymmetry is written as follows: 
\begin{eqnarray}
  \Amue=\Amuevac+\hat A \Amuemat +\mathcal{O}(\hat A^2) \,,
\end{eqnarray}
where
\begin{equation}
  \Amuevac = \frac{-28 \alpha\Delta\cos\theta_{23}\sin\delta_{\mathrm{CP}}\sin\Delta }{3\sqrt{2}\sin\theta_{23}\sin\Delta+28\alpha\Delta \cos\theta_{23}\cos\delta_{\mathrm{CP}}\cos\Delta}\label{eq:Avac}
\end{equation} 
\begin{equation}  
\resizebox{.9\hsize}{!}{ $\Amuemat = -\sin^2\theta_{23}(\Delta\cos\Delta-\sin\Delta)\frac{126\alpha\Delta\cos\theta_{23}\cos\delta_{\mathrm{CP}}\cos\Delta+18\sin^2\theta_{23}\sin\Delta}{(3\sin^2\theta_{23}\sin\Delta+7\sqrt{2}\alpha\cos\delta_{\mathrm{CP}}\cos\Delta\sin^2(2\theta_{23}))^2}$}
\label{Amat} 
\end{equation} 
It is clear that when the value of $\theta_{23}$ increases, the denominator of both the contributing term in Eq.~\ref{eq:Avac} and Eq.~\ref{Amat} increases. For this reason, the absolute value of the asymmetry becomes smaller, and we expect less CPV sensitivity. So, at the first oscillation maximum, ($\Delta=\pi/2$)\footnote{To be maximally sensitive to the oscillation probability, we must have $\Delta = (2n + 1) \frac{\pi}{2}$, where $n = 0, 1, 2,...$} the asymmetry reduces to:
\begin{eqnarray}
  \Amue&\approx &-\frac{7}{3}\alpha\sqrt{2}\pi\cot\theta_{23}\sin\delta_{\mathrm{CP}}+2\hat A
  \label{FOMmue}\,,
\end{eqnarray}
whose modulus decreases with an increase in $\theta_{23}$. The genuine or intrinsic CP has a sign opposite to the extrinsic or fake matter-induced contribution. Thus, there exists some combination of $\theta_{23}$ and $\dcp \in [0^{\circ},180^{\circ}]$, such that this asymmetry vanishes. It is interesting to note that the vacuum contribution becomes three times larger when considering the second oscillation maximum ($\Delta=3\pi/2$). Therefore, observing the CPV at such $L/E$ combination can give much more sensitivity to  $\delta_{\mathrm{CP}}$ \cite{Rout:2020emr,Blennow:2019bvl,Tang:2019wsv}. The exact numerical behavior of the CP asymmetry in the appearance channel ($|\Amue|$) is shown in Fig.~\ref{fig:1} for ($L$ = 1285 km, $E$ = 2.5 GeV), ($L$ = 1285 km, $E$ = 0.9 GeV)  which corresponds to the first and second oscillation maxima in DUNE. For each $L/E$ combination, we show three panels: in the left column, we show the vacuum or $\delta_{\mathrm{CP}}$-induced contribution (intrinsic). In the center, we illustrate the total asymmetry, and in the right column, we display the contribution due to the matter effects (extrinsic). In all the panels, we only plot the absolute value of the asymmetries since, in this work, the most important aspect is to stress on the difference between the asymmetries in the CP-violating and the CP-conserving cases. From the top left panel in Fig.~\ref{fig:1}, we observe that the intrinsic contribution is the same in both the maximal CP-violating values of $\delta_{\mathrm{CP}}$ (90 and -90$^\circ$). Moreover, keeping the CP phase fixed to any value, the asymmetry reduces when we increase the value of $\theta_{23}$ from lower octant\footnote{Existence of non-maximal $\theta_{23}$ gives rise to two degenerate solutions: $\theta_{23} < 45^{\circ}$ (LO) and $\theta_{23} > 45^{\circ}$ (HO).} (LO) to higher octant (HO) (as expected from Eq.~\ref{eq:Avac} and Eq.~\ref{Amat}). Contrastingly, the extrinsic CP asymmetry (top right panel), which occurs solely due to the matter effect, is asymmetric, being larger for favorable choices of $\delta_{\mathrm{CP}}$ in NMO (choices of $\dcp$ which enhances $\nu_{\mu}\rightarrow \nu_{e}$ in the presence of matter, corresponding to the negative half plane, \ie~ $\dcp \in [180^{\circ}, 360^{\circ}]$) and smaller for unfavorable choices of $\delta_{\mathrm{CP}}$ in NMO (positive half plane, \ie~ $\dcp \in [0^{\circ},180^{\circ}]$). Therefore, the total asymmetry (middle panel) is no longer the same for the maximal CP-violating values of $\delta_{\mathrm{CP}}$. Further, due to contribution from $\hat A$, the intrinsic $|\Amue|$, which is zero at CP-conserving values ($\delta_{\mathrm{CP}} = 0^\circ\, \text{and}\,180^\circ$), now has a finite value. Hence, from the top middle panel, it is clear that the asymmetries in CP-violating cases tend to shift closer to the CP-conserving value when $\theta_{23}$ increases. For the bottom row, where we plot the CP asymmetry at the second oscillation maxima ($E$ = 0.9 GeV), the matter effect becomes less, and the intrinsic component completely dominates the total asymmetry. Moreover, the asymmetry values are amplified in the bottom panel, which we expect as here the $L$ and $E$ are similar to the second oscillation maximum in DUNE.
Similarly, in Fig.~\ref{fig:2}, we plot the CP asymmetry for the T2HK (JD) setup with $L=295$ km and $E=0.6$ GeV at the first oscillation maxima (top row) and T2HKK with $L=1100$ km and $E=0.6$ GeV at the second oscillation maxima (bottom row).
The top panel does not observe any significant contribution in the extrinsic panel due to less matter effect ($L$ = 295 km). Thus, we expect the J-PARC based experiments to provide a cleaner environment for the measurements in $\delta_{\mathrm{CP}}$, even though the values reached by the asymmetries in these cases are not as high as the DUNE. On the other hand, the bottom panels of Fig.~\ref{fig:2}, which correspond to the second oscillation maximum $L/E$ choice for T2HKK, behave just like the bottom panels in Fig.~\ref{fig:1}.

\subsection{Extrinsic CP Asymmetry in Disappearance Channel}
\label{sec:2b}

\begin{figure}[htb!]
     \centering
     \includegraphics[scale=1]{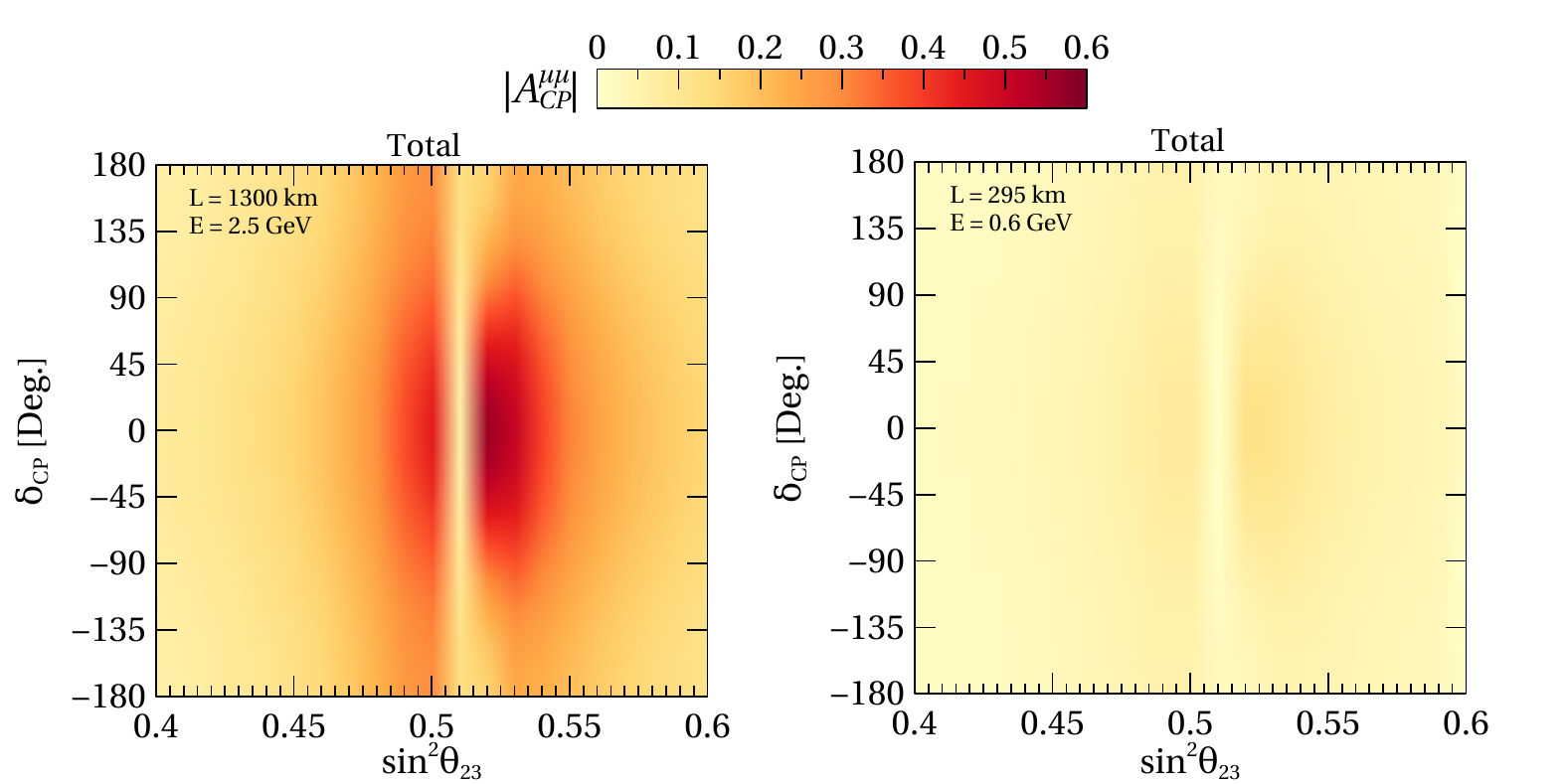}
     \mycaption{ $|\mathcal{A}^{\mu \mu}_{\mathrm{CP}}|$ as a function of $\delta_{\mathrm{CP}}$ and $\sin^{2}\theta_{23}$ assuming NMO for first oscillation minimum in DUNE ($L$ = 1285 km, $E$ = 2.5 GeV) and JD ($L$ = 295 km, $E$ = 0.6 GeV) is shown in the left and right panel, respectively. Values of other oscillation parameters are taken from Table~\ref{table:one}.} 
     \label{fig:3}
 \end{figure}
The $\nu_{\mu}$ disappearance channel, being T invariant, is directly CP-conserving in a vacuum-like scenario. However, the Earth matter potential interacts differently with neutrinos and antineutrinos, which can further induce a fake or extrinsic CPV in this channel \cite{Giarnetti:2021wur,Ohlsson:2014cha}. Thus, although the disappearance channel does not directly affect the measurements in CP phase, it is important to realize its ability to generate fake or extrinsic CPV. This is crucial in our study as we will discuss later its effect in our results (Sec.~\ref{sec:4a}).
Following the same convention, as discussed in Ref.~\cite{Agarwalla:2021bzs}, we write the disappearance probability as
\begin{eqnarray}
  P_{\mu\mu}\approx 1-M\sin^2(2\theta_{23})-N\sin^2\theta_{23}-R\sin2\theta_{23}+T\sin4\theta_{23}\,,
\end{eqnarray}
where:
\begin{eqnarray}
 \nonumber M&=&\sin^2\Delta-\alpha\cos^2\theta_{12}\Delta\sin2\Delta+ \\
  & &+\frac{2}{\hat A-1}\sin^2\theta_{13}\left(\sin\Delta\cos(\hat A\Delta)\frac{\sin[(\hat A-1)\Delta]}{\hat A-1}-\frac{\hat A}{2}\Delta \sin2\Delta\right)\,,\label{eq:disapp_M} \\
  R&=&2\alpha\sin\theta_{13}\sin2\theta_{12}\cos\delta_{\mathrm{CP}}\cos\Delta\frac{\sin\hat A\Delta}{\hat A}\frac{\sin[(\hat A-1)\Delta]}{\hat A-1} \,, \label{eq:disapp_R}\\
  T&=&\frac{1}{\hat A-1}\alpha \sin\theta_{13}\sin2\theta_{12}\cos\delta_{\mathrm{CP}}\sin\Delta\left(\hat A \sin\Delta -\frac{\sin\hat A \Delta}{\hat A}\cos[(\hat A-1)\Delta]\right)\,.\label{eq:disapp_T}\nonumber \\ && \,
\end{eqnarray}
and N has already been defined in Eq. \ref{eq:2}. The detailed analytical discussion of fake CP asymmetry in the disappearance channel results in a cumbersome expression. However, for first oscillation minima ($\Delta=\pi/2$) and the approximated numerical values of the solar and the reactor mixing angles ($\sin\theta_{12}=1/\sqrt{3}$ and $\sin\theta_{13}=1/7$), one can calculate the CP asymmetry in the $\nu_\mu\rightarrow\nu_\mu$ disappearance channel which is defined as
\begin{equation}
\label{eq:CPA_disapp}
\mathcal{A}^{\mu\mu}_{\mathrm{CP}}=\dfrac{P_{\mu \mu}-\bar P_{\mu \mu}}{P_{\mu \mu}+\bar P_{\mu \mu}}\,
\end{equation}
Substituting the discussed approximation in the above expression and neglecting the higher order terms, we get
\begin{eqnarray}
 \mathcal{A}^{\mu\mu}_{\mathrm{CP}}\approx\hat{A}\frac{24\sin^2\theta_{23}+7\sqrt{2}(\pi^2-4)\alpha\cos\delta_{\mathrm{CP}}\sin2\theta_{23}}{6+141\cos2\theta_{23}}\,.
 \label{eq:4}
\end{eqnarray}
This asymmetry increases with the increase in $\theta_{23}$ until the expansion breaks at $\cos2\theta_{23}= - 6/141$. This occurs for $\sin^{2}\theta_{23} > 0.5$ (HO). While after this value, the magnitude of asymmetry starts decreasing with the increase in $\theta_{23}$. In Fig.~\ref{fig:3}, we exhibit the absolute value of the disappearance asymmetry ($|\mathcal{A}^{\mu\mu}_{\mathrm{CP}}|$) for ($L$ = 1285 km, $E$ = 2.5 GeV) and ($L$ = 295 km, $E$ = 0.6 GeV) that also corresponds to DUNE and JD at their respective first oscillation maxima energy. We do not show the plots corresponding to the second oscillation maxima since the fake CP asymmetry in disappearance is solely due to the interaction with the Earth matter potential, whose effect becomes minimal in such conditions. Thus, we do not expect $\mathcal{A}^{\mu\mu}_{\mathrm{CP}}$ in DUNE and KD working at their second oscillation maxima to have any significant contribution to our analysis. We notice that JD, bearing a relatively smaller baseline ($L$ = 295 km), has very small matter effects and thus exhibits very minute fake asymmetry even at first oscillation maximum, as shown in the right panel of Fig.~\ref{fig:3}.  On the other hand, DUNE has a larger baseline ($L$ = 1285 km), thus exhibiting consequential $|\mathcal{A}^{\mu\mu}_{\mathrm{CP}}|$ that reaches as high as $\approx 0.6$ (see the left panel in Fig.~\ref{fig:3}), which is almost comparable to the total $|\mathcal{A}^{\mu e}_{\mathrm{CP}}|$ ($\approx$ 0.8 ) (see the top middle panel in Fig.~\ref{fig:1}). We observe that $|\mathcal{A}^{\mu\mu}_{\mathrm{CP}}|$ is minimal at the two extremes of octant of $\theta_{23}$ for any value of $\delta_{\mathrm{CP}}$. The asymmetry gradually increases while proceeding towards the maximal mixing (MM) corresponding to $\sin^{2}\theta_{23}=0.5$, from either side for almost all $\delta_{\mathrm{CP}}$. However, $|\mathcal{A}^{\mu\mu}_{\mathrm{CP}}|$ manifests two maxima around $\delta_{\mathrm{CP}} = 0^{\circ}$, one each in the two octants: LO ($\sin^2\theta_{23} \approx 0.49$) and HO ($\sin^2\theta_{23} \approx 0.52$). As discussed previously in the analysis of Eq.~\ref{eq:4}, we observe a critical point in HO in the figure as well, around which the nature of $\mathcal{A}^{\mu\mu}_{\mathrm{CP}}$ changes. Correspondingly, around this point, our analytical expansion also breaks. However, in our expression, this completely vanishes ($\mathcal{A}^{\mu\mu}_{\mathrm{CP}} \approx 0$) as we have neglected the higher order terms. This nature of fake $\mathcal{A}^{\mu\mu}_{\mathrm{CP}}$ is crucial in our result, as we will elaborate on this further in our results (Sec.~\ref{sec:4a}).
%
\subsection{$\theta_{23}-\delta_{\mathrm{CP}}$ Degeneracy}
\label{sec:2c}
%
From the above discussion, we observe that the value of the atmospheric mixing angle can influence CPV sensitivity. We can further expect this sensitivity to be affected by the octant of $\theta_{23}-\delta_{\mathrm{CP}}$ degeneracy in appearance channel~\cite{Agarwalla:2013ju,Minakata:2013eoa,Coloma:2012wq}. The persisting issue of octant of $\theta_{23}$~\cite{Fogli:1996pv} makes it highly probable for some $\bar\theta_{23}$ and $\bar\delta_{\mathrm{CP}}$ to exist such that for a given $\theta_{23}$, $\delta_{\mathrm{CP}}$,
\begin{eqnarray}
 P_{\mu e}(\theta_{23},\delta_{\mathrm{CP}})&=&P_{\mu e}(\bar\theta_{23},\bar\delta_\mathrm{{CP}}) \,,
 \label{systema}\\
 \bar P_{\mu e}(\theta_{23},\delta_{\mathrm{CP}})&=&\bar P_{\mu e}(\bar\theta_{23},\bar\delta_{\mathrm{CP}})\,,
 \label{systemb}
\end{eqnarray}
holds true.
 For instance, fixing $\theta_{23}=45^\circ$, $\sin\theta_{13}=1/7$, and $\sin\theta_{12}=1/\sqrt{3}$ in presence of matter effect, $\nu_{\mu}\to \nu_{e}$ oscillation probability is given as:
\begin{eqnarray}
 P_{\mu e}(45^{\circ},\delta_{\mathrm{CP}})=\frac{2\sin[(\hat{A}-1)\Delta]^2}{49(\hat{A}-1)^2}+\frac{4\sqrt{2}\alpha\cos(\delta_{\mathrm{CP}}+\Delta)\sin[(\hat{A}-1)\Delta]\sin(\hat{A}\Delta)}{21 \hat{A}(\hat{A}-1)}\,.
\end{eqnarray}
Assuming that the degenerate atmospheric angle is not too far from the maximal value, we define it as $\bar\theta_{23}= 45^{\circ}+x$. Now, considering terms only up to the first order in $x$ we obtain:
\begin{eqnarray}
 P_{\mu e}(45^{\circ}+x,\bar\delta_{\mathrm{CP}})=P_{\mu e}(45^{\circ},\bar\delta_{\mathrm{CP}})+\frac{4x \sin[(\hat{A}-1)\Delta]^2}{49(\hat{A}-1)^2}\,.
\end{eqnarray}
For the above scenario, the system of equations in Eq.~\ref{systema} and Eq.~\ref{systemb}
reduces to the following two equations-
\begin{subequations}
\begin{equation}
 \frac{\sqrt{2}\alpha\sin(\hat{A}\Delta)}{3\hat{A}}[\cos(\Delta-\delta_{\mathrm{CP}})-\cos(\Delta-\bar\delta_{\mathrm{CP}})] = x\frac{\sin[(\hat{A}-1)\Delta]}{7(\hat{A}-1)}\, \mathrm{and}
  \label{rhsa}
 \end{equation}
  \begin{equation}
 \frac{\cos(\Delta+\delta_{\mathrm{CP}})-\cos(\Delta+\bar\delta_{\mathrm{CP}})}{\cos(\Delta-\delta_{\mathrm{CP}})-\cos(\Delta-\bar\delta_{\mathrm{CP}})} =  \frac{\sin[(\hat{A}-1)\Delta]}{\sin[(\hat{A}+1)\Delta]}\frac{1+\hat{A}}{\hat{A}-1}\,.
 \label{rhsb}
\end{equation}
\end{subequations}
So, for each value of $\Delta$, the above two equations will have different solutions. Thus, in principle, spectral analysis can reduce the above-mentioned degeneracy. Then, it is clear that in the vacuum-like scenario ($\hat{A}\to0)$, Eq.~\ref{rhsa} gives $x=0$, while in Eq.~\ref{rhsb} $\bar\delta_{\mathrm{CP}}=\delta_{\mathrm{CP}}$. Therefore in an experiment with negligible matter effect, the role of this degeneracy will not be crucial in determining the sensitivity of $\delta_{\mathrm{\mathrm{CP}}}$. While in the presence of substantial matter effect and taking as an example $\delta_{\mathrm{CP}}=90^\circ$, the above equations will always have a solution for $\bar\delta_{\mathrm{CP}}$ except when $\Delta= \pi/2$. Moreover, we have checked that given a certain matter potential, there is always a value of $\Delta < \pi/2$,   for which we obtain $\bar\delta_{\mathrm{CP}}=0$. This infers that, in the presence of matter effect, we can always have identical solutions for maximal CP-violating and a CP-conserving case. The corresponding value of $x$ in such cases is always positive, implying that the degenerate solution lies in the higher octant. While on the other hand, when $\delta_{\mathrm{CP}}=-90^\circ$, we obtain degenerate solutions for $\bar\delta_{\mathrm{CP}}$, but to get the CP-conserving degenerate solution, we need to find a specific $L/E$ ratio that exceeds the value which we obtain at the atmospheric peak. Moreover, in this case, the corresponding deviation of the mixing angle is negative; hence the degenerate solution lies in the lower octant. 

Next, we discuss the $\theta_{23}-\delta_{\mathrm{CP}}$ degeneracy when considering both disappearance and appearance channels. For the degeneracy to occur with all other mixing angles and mass-splittings kept fixed, there should exist $\bar\theta_{23}$ and $\bar\delta_{\mathrm{CP}}$ such that
\begin{eqnarray}
 P_{\mu \mu}(\theta_{23},\delta_{\mathrm{CP}})&=&P_{\mu \mu}(\bar\theta_{23},\bar\delta_{\mathrm{CP}}) \, \\
 \bar P_{\mu \mu}(\theta_{23},\delta_{\mathrm{CP}})&=&\bar P_{\mu \mu}(\bar\theta_{23},\bar\delta_{\mathrm{CP}}).
\end{eqnarray}
However, the disappearance probability only has a mild dependence on $\delta_{\mathrm{CP}}$, which plays an important role when there is a substantial matter effect and $\theta_{23}$ around maximal mixing, as elaborated previously in the discussion of Fig.~\ref{fig:3}. Otherwise, the disappearance channel precisely measures the atmospheric mixing angle in $\delta_{\mathrm{CP}}$ independent way. Thus disappearance channel helps in constraining $\theta_{23}$ when it is not around maximal mixing, which reduces the effect of $\theta_{23}-\delta_{\mathrm{CP}}$ degeneracy for these $\theta_{23}$ in the appearance channel as well. Thus a combined analysis between the appearance and disappearance channels is expected to break the $\theta_{23}-\delta_{\mathrm{CP}}$ degeneracy only if (a) the matter effects are negligible or (b) true values of the atmospheric mixing angle are far away from the maximal mixing scenario, which for DUNE is true if $\sin^2\theta_{23}\notin[0.48 , 0.55]$ (refer to the discussion around Fig.~\ref{fig:3}). Therefore, in this region we obtain a  $\theta_{23}$ independent $\delta_{\mathrm{CP}}$ measurement.  
%
\section{ Experimental Features and Simulation Details}
\label{sec:3}
%
\begin{table}[htb!]
\resizebox{\columnwidth}{!}{%
    \centering
    \begin{tabular}{|c|c|c|}
    \hline \hline
       Characteristics  & DUNE & JD/KD  \\
       \hline \hline
       Baseline (km) & 1285 & 295 (1100)\\
       \hline
       $\rho_{\mathrm{avg}}$ (g/cm$^{3}$) & 2.848 & 2.7 (2.8)\\
       \hline
       Beam & LBNF~\cite{DUNE:2020lwj} & J-PARC~\cite{Abe:2018uyc}\\
       \hline
       Beam Type & wide-band, on-axis & narrow-band, 2.5$^{\circ}$ off-axis\\
       \hline
       Beam Power & 1.2 MW & 1.3 MW\\
       \hline
       Proton Energy & 120 GeV & 30 GeV\\
       \hline
       P.O.T./year & 1.1 $\times$ 10$^{21}$ & 2.7 $\times$ 10$^{22}$\\
       \hline
       Flux peaks at (GeV) & 2.5  & 0.6  \\
       \hline
       1$^{\mathrm{st}}$ ( 2$^{\mathrm{nd}}$) oscillation maxima & \multirow{2}{*}{2.6 (0.87) }& \multirow{2}{*}{0.6 (0.2) / 1.8 (0.6)}  \\
       for appearance channel (GeV) & &\\
       \hline
       Detector mass (kt) & 40, LArTPC & 187 each, water Cherenkov\\
       \hline
       Runtime ($\nu + \bar{\nu}$) yrs & 5 + 5 & 2.5 + 7.5\\
       \hline
       Exposure (kt$\cdot$MW$\cdot$yrs)  & 480 & 2431\\
       \hline
       Signal Norm. Error (App.) & 2\% & 5\%\\
       \hline
       Signal Norm. Error (Disapp.) & 5\% & 3.5\%\\
       \hline
       Binned-events & \multirow{2}{*}{~\cite{DUNE:2021cuw} }& \multirow{2}{*}{~\cite{Hyper-Kamiokande:2016srs}} \\
       matched with & &\\
       \hline \hline
    \end{tabular}}
    \mycaption{Essential experimental features of various long-baseline experiments considered in our analysis. The first column characterizes DUNE, while the second column depicts aspects of JD and KD.}
    \label{table:two}
\end{table}

The LBL accelerator experiments allow us to probe neutrino oscillation 
phenomena in a very controlled environment with fixed baselines and well-known
energy spectra of neutrinos~\cite{Feldman_2013,Agarwalla:2014fva,Diwan_2016,Giganti:2017fhf}. 
Over the last two decades or so, various LBL experiments have contributed significantly 
in measuring the three-flavor oscillation parameters with high precision~\cite{Capozzi:2021fjo,ParticleDataGroup:2022pth}.
These experiments have access to both $\nu_\mu \to \nu_e$ appearance and $\nu_\mu \to \nu_\mu$ 
disappearance channels with a very well-determined initial $muon$ neutrino flux. The upcoming next-generation LBL experiments are expected to resolve all the degeneracies~\cite{Barger:2001yr,Huber:2003ak,Donini:2003vz,Agarwalla:2008jin} among the three-flavor neutrino oscillation parameters with a very high confidence level and improve the existing precision manifold.

In our analysis, we consider the two upcoming next-generation LBL experiments, DUNE and T2HK (JD). We also take into account the possibility of the second detector in Korea, the T2HKK (JD + KD) setup. The unprecedented high statistics, intense beam, and reduced systematic uncertainties in these experiments will be able to enhance precision measurements in oscillation parameters and possibly open the door for studying New Physics in the lepton sector. DUNE is proposed to bear a long baseline (1285 km) and so will be largely influenced by the matter effect, while JD, having a relatively much smaller baseline (295 km), will portray an almost vacuum-like scenario. Also, the flux in DUNE will peak at higher energies ($\sim2.5$ GeV), which enables the potential for searching for $\nu_\tau$ appearance as well \cite{Machado:2015vwa,Ghoshal:2019pab, DeGouvea:2019kea,Machado:2020yxl,MammenAbraham:2022xoc}, while the flux in JD will peak around $\sim 0.6$ GeV. Further, the DUNE far-detector will observe a wide-band on-axis beam, which allows for scanning several different $L/E$ ratios (including the second oscillation maximum, which occurs at 0.9 GeV), while the JD detector will be receiving a narrow-band off-axis beam. In the JD + KD setup, the proposed second detector, KD in Korea (receiving the same flux as JD with a baseline of $L$ = 1100 km), is expected to work at the second oscillation maximum. The DUNE will be a 40 kt LArTPC, with great imaging capabilities that will lower signal normalization uncertainties to just 2\% in appearance and 5\% in disappearance. On the other hand, JD and KD, each will be 187 kt water Cherenkov detectors, which will accumulate huge statistics, however, with relatively poorer expected signal normalization uncertainties of 5\% in the appearance channel and 3.5\% in the disappearance. Further, the sources of backgrounds and their contribution to normalization uncertainties have been taken from Ref.~\cite{DUNE:2021cuw,Hyper-Kamiokande:2016srs}. The proposed runtime ratio in JD/KD is in contrast with DUNE. While DUNE proposes a balanced run in neutrino and antineutrino mode, T2HK will be running in the ratio of 1:3 in $\nu : \bar{\nu}$ mode to ensure similar statistics from both modes. In Table~\ref{table:two}\,, we enlist characteristic attributes in these next-generation experiments that sum up their complementary features.
\begin{table}
\resizebox{\columnwidth}{!}{%
\centering
\begin{tabular}{ |c|c|c|c|c|}
\hline
\multirow{2}{*}{\diagbox[width=10em]{Experiment}{Parameter}} &\multirow{2}{*}{$\theta_{23}$}&$\delta_{\mathrm{CP}}$= 0$^{\circ}$ & $\delta_{\mathrm{CP}}=90^{\circ}$ & $\delta_{\mathrm{CP}}=-90^{\circ}$ \\
&&($\nu_e$, $\bar{\nu}_e$, $\mathcal{N}^{\mu e}_{\mathrm{CP}}$)&($\nu_e$, $\bar{\nu}_e$, $\mathcal{N}^{\mu e}_{\mathrm{CP}}$)&($\nu_e$, $\bar{\nu}_e$, $\mathcal{N}^{\mu e}_{\mathrm{CP}}$)\\
\hline
\multirow{3}{4em}{DUNE} &
$40^{\circ}$ & 1965, 812, \it{0.41} &1657, 857, \it{0.31} &2303, 728, \it{0.52}\\
& $45^{\circ}$ & 2215, 875, \it{0.43} &1902, 920, \it{0.34} & 2558, 790, \it{0.53}\\
& $50^{\circ}$ & 2470, 938, \it{0.45} &2161, 982, \it{0.37} & 2807, 854, \it{0.53} \\
\hline
\multirow{3}{4em}{ JD} & $40^{\circ}$ & 1644, 1420, \it{0.074} & 1277, 1687, \it{-0.14}  & 2024, 1113, \it{0.29}\\
& $45^{\circ}$ & 1890,1594, \it{0.085}  & 1517, 1868, \it{-0.1}  & 2276, 1286, \it{0.28} \\
& $50^{\circ}$ &  2137, 1770, \it{0.093} & 1770, 2041, \it{-0.07}  & 2518, 1476, \it{0.26}\\
\hline
\end{tabular}}
\mycaption{
Total appearance event rates (signal + background) in neutrino, antineutrino modes, and 
$\mathcal{N}^{\mu e}_{\mathrm{CP}}$ for DUNE (JD) corresponding to different sets of 
$\delta_{\mathrm{CP}}: 0^{\circ}, 90^{\circ}, -90^{\circ}$ and $\theta_{23} : 40^{\circ}, 45^{\circ}$, 
and 50$^{\circ}$ are shown in the second (third) set of rows, respectively.
}
\label{table:three}

\end{table}

In Table~\ref{table:three}\,, we summarize the number of neutrino and antineutrino events for DUNE and JD for three different choices of $\delta_{\mathrm{CP}}$ ($0^\circ,\,90^\circ,\,-90^\circ)$ and three choices of $\theta_{23}$: LO ($40^\circ$), MM ($45^\circ$), and in the HO ($50^\circ$). For the sake of comparison, we also compute the values of the integrated asymmetries, defined as
\begin{eqnarray}
 \mathcal{N}^{\mu e}_{\mathrm{CP}}=\frac{N_{\mu e}-\bar{N}_{\mu e}}{N_{\mu e}+\bar{N}_{\mu e}}\,,
\end{eqnarray}
where $N_{\mu e}$ ($\bar N_{\mu e}$) is the number of events in the neutrino (antineutrino) mode. JD statistics are higher than DUNE, following the higher exposure in the J-PARC (2431 kt$\cdot$MW$\cdot$yrs) than in the Fermilab-based experiment (480 kt$\cdot$MW$\cdot$yrs). Moreover, as expected, for the favorable choice of parameters, i.e., NMO, $\delta_{\mathrm{CP}}=-90^\circ$ ($\delta_{\mathrm{CP}}= 90^\circ$), and HO, we observe the highest neutrino (antineutrino) events. The integrated asymmetries follow the same nature as we observe at the probability level (refer to Fig.~\ref{fig:1}). Here also $\mathcal{N}^{\mu e}_{\mathrm{CP}}$ in the CP-violating case tends to come closer to its value in the CP-conserving case as $\theta_{23}$ increases in both the experiments.
%
\section{Our Results}
\label{sec:4}

In this section, we discuss the abilities of DUNE, T2HK (JD), and their combination in achieving the landmark of 75\% CP coverage in true $\dcp$ for leptonic CPV at 3$\sigma$ C.L. for all the values of $\sin^{2}\theta_{23}$ in Nature. We also explore the capability of the T2HKK (JD + KD) setup. Further, we also inspect the effect of changing overall exposure in each setup while determining the CP coverage. Next, we survey the optimal runtime in neutrino and antineutrino modes to effectively increase CP coverage in different experimental setups. Finally, we address the importance of systematic uncertainties and how their variation can significantly affect our results.

In our analysis, we define the CP coverage as the percentage of true $\delta_{\mathrm{CP}}$ for which an experiment establishes at least a 3$\sigma$ CPV sensitivity. For this, we generate our prospective data assuming the entire range of true $\delta_{\mathrm{CP}} \in [-180^{\circ} , 180^{\circ}]$, while in the fit, we marginalize over the test $\delta_{\mathrm{CP}} = 0^{\circ}$ and $180^{\circ}$ and choose the minimum. We use the Poissonian $\chi^2$~\cite{Baker:1983tu} and estimate the median sensitivity~\cite{Cowan:2010js} of a given LBL experiment in the frequentist approach~\cite{Blennow:2013oma}. To evaluate the sensitivity towards leptonic CPV, we use the following definition of $\Delta\chi^2$:
\begin{equation}
	\Delta \chi^2 = \underset{(\delta^{\mathrm{test}}_{\mathrm{CP}}, \,\theta_{23},\,\lambda_1,\, \lambda_2)}{\mathrm{min}} \,\bigg[\chi^2(\delta^{\mathrm{true}}_{\mathrm{CP}})-\chi^2(\delta^{\mathrm{test}}_{\mathrm{CP}}=0^{\circ}\, \text{and}\, 180^{\circ})\bigg]\,.
 \label{eq:chi2-cpv}
\end{equation}
The fit is performed by marginalizing over $\theta_{23}$ in its current 3$\sigma$ range of [0.4 , 0.6] while keeping
all other parameters fixed at the benchmark values as mentioned in Table~\ref{table:one}, assuming NMO. 
The symbols $\lambda_{1}$ and $\lambda_2$ refer to the systematic pulls~\cite{Huber:2002mx,Fogli:2002pt,Gonzalez-Garcia:2004pka,Ankowski:2016jdd} on signal and background, respectively. The present global neutrino oscillation data~\cite{Capozzi:2021fjo,NuFIT,deSalas:2020pgw,Esteban:2020cvm} measures atmospheric mass splitting with a very high precision of 1.1\%, which will further improve to 0.5\% with six years of data taking by JUNO~\cite{JUNO:2022mxj}. Therefore, we do not marginalize over the present uncertainty in $\Delta m^{2}_{31}$. Also, we do not marginalize over the wrong mass ordering in any of the results as there are hints towards NMO from the global oscillation data~\cite{Capozzi:2021fjo,deSalas:2020pgw,NuFIT,Esteban:2020cvm}. Moreover, the currently running LBL experiments: NO$\nu$A and T2K, and the atmospheric experiments: Super-K~\cite{FernandezMenendez:2021jfk} and DeepCore~\cite{LeonardDeHolton:2022evl} will further strengthen the mass ordering measurements in the near future. Also, by the time DUNE accumulates data for CPV searches, it is expected to already fix the mass ordering~\cite{DUNE:2020ypp}. For all the simulations, we use the GLoBES software \cite{Huber:2004ka,Huber:2007ji}. 
%
\subsection{Impact of $\theta_{23}$ on CP coverage}
\label{sec:4a}
 \begin{figure}[htb!]
     \centering
     \includegraphics[width=0.8\linewidth]{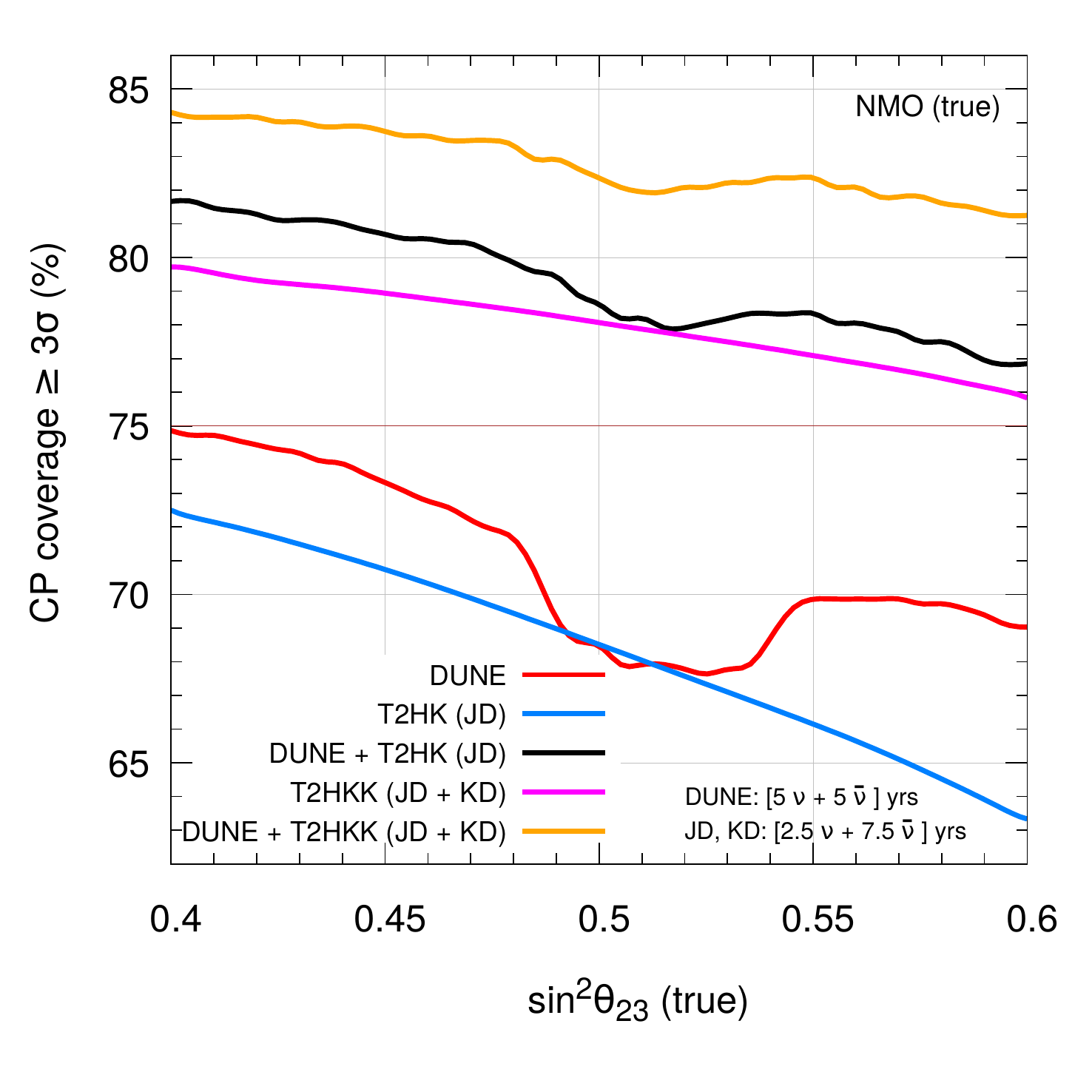}
     \mycaption{Coverage in true $\dcp$ for achieving $\geq 3 \sigma$ leptonic CPV as a function of true $\sin^{2}\theta_{23}$, when marginalized over the current 3$\sigma$ uncertain range of $\sin^{2}\theta_{23}$ [0.4 , 0.6] in the theory. The curves: red, blue, black, magenta, and orange are for DUNE, JD, DUNE + JD, JD + KD, and DUNE + JD + KD neutrino oscillation experiments, respectively. We assume true NMO, benchmark exposure, and the nominal runtime as mentioned in Table~\ref{table:two} in both data and theory.   }
 \label{fig:4}
 \end{figure}
 \begin{figure}[htb!]
     \centering
     \includegraphics[width=0.8\linewidth]{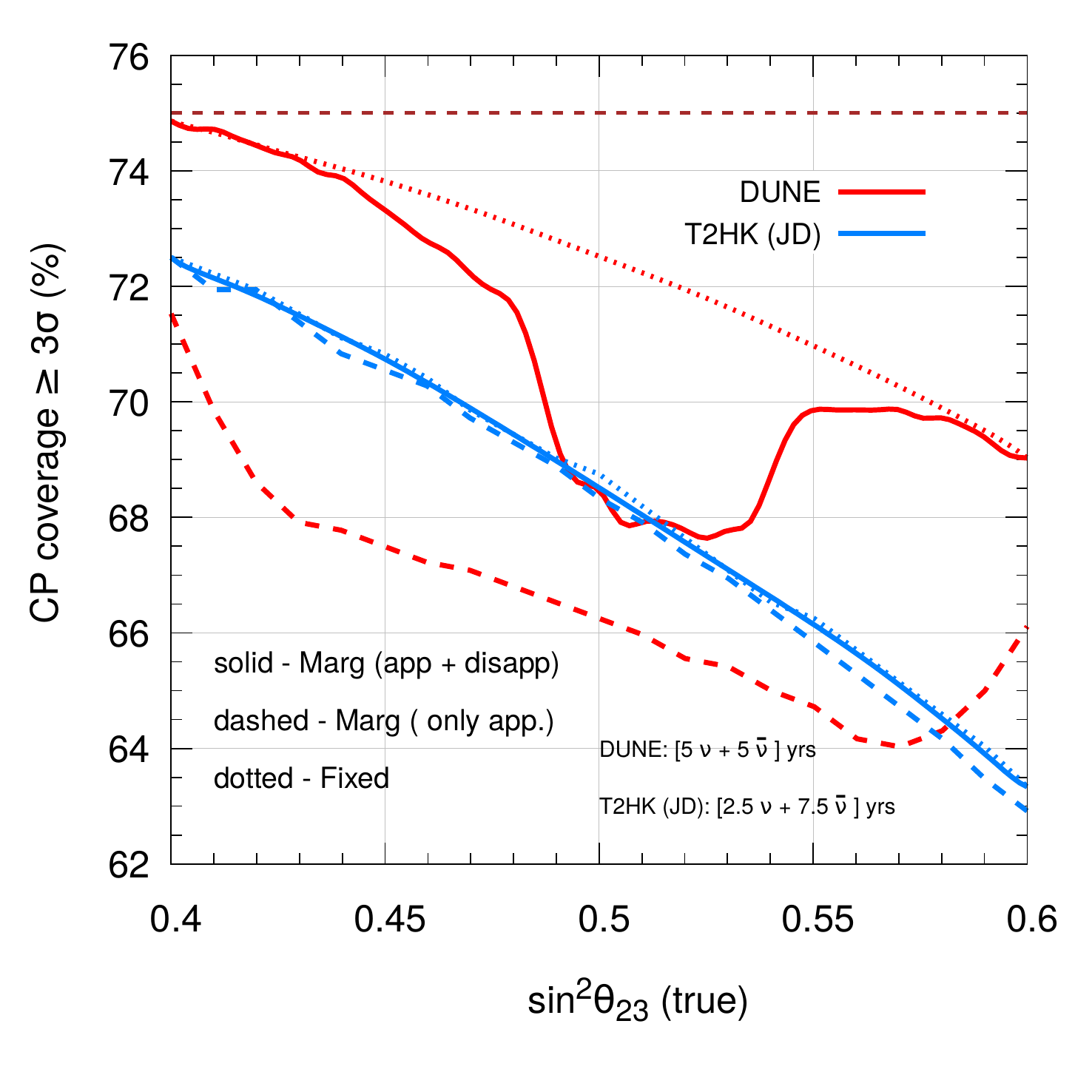}
     \mycaption{Coverage in true $\dcp$ for achieving $\geq 3 \sigma$ leptonic CPV as a function of true $\sin^{2}\theta_{23}$ assuming NMO. The solid curves are our result from Fig.~\ref{fig:4}, in which we generate data with both appearance and disappearance channels and marginalize over $\sin^{2}\theta_{23}$ in its 3$\sigma$ uncertain range of [0.4 , 0.6] in the fit. The dashed curves are obtained by generating our data with only appearance channel and marginalizing over $\sin^{2}\theta_{23}$ in its 3$\sigma$ uncertain range in the fit. While the dotted curves correspond to our result by fixing identical $\sin^{2}\theta_{23}$ in both data and fit (fixed-parameter scenario). The curves: red and blue, correspond to DUNE and JD, respectively.}
     \label{fig:5}
\end{figure}
As previously discussed in Sec.~\ref{sec:2}, the atmospheric mixing angle can play an important role in determining the CPV sensitivity. In Fig.~\ref{fig:4}, we depict how the ability of an experimental setup to measure CP coverage changes with the true value of $\sin^{2}\theta_{23}$. We generate our data for each $\sin^{2}\theta_{23}$ (true) by varying $\sin^{2}\theta_{23}$ in our theory throughout the uncertain range of [0.4 , 0.6]. Each colored curve corresponds to the CP coverage of a particular setup as a function of the true value of $\sin^2\tzm$. We observe that none of the individual experiments: DUNE (red curve) or JD (blue curve) can achieve the milestone of 75\%. However, their combination makes CP coverage for the entire canvas of $\sin^{2}\theta_{23}$ above 75\%. This points out that the complementarity between DUNE and JD or JD and KD can help attain a better CP coverage irrespective of the $\sin^{2}\theta_{23}$ value in Nature. 
 
 Comparatively, DUNE + T2HK has consistently better CP coverage than T2HKK. However, we would like to underline that T2HK could still be benefitted from the addition of a second Korean detector when its data is combined with  both DUNE + T2HK (refer to the orange-colored curve in Fig.~\ref{fig:4}). Working at second oscillation maxima, KD has the advantage of being more sensitive to $\dcp$, getting less affected by the matter potential. However, since the same beam is proposed to be shared between JD and KD, the flux being $\propto 1/L^2$, KD receives a substantially less flux. Thus, KD will be unable to unlock the full potential of the second oscillation maxima. 
 
In all the setups, there is a general notion of CP coverage decreasing as we increase $\sin^{2}\theta_{23}$ in the data; the reason for that can be found in the behavior of the appearance asymmetry, as previously discussed in text around figures~\ref{fig:1} and~\ref{fig:2}. In DUNE, the $3\sigma$ CP coverage decreases from 75\% to 68\% when $\theta_{23}$ increases from $40^\circ$ ($\sin^2\theta_{23}=0.4$) to $60^\circ$ ($\sin^2\theta_{23}=0.6$), while in JD the coverage decreases from 73\% to 63\% of true $\dcp$ for the same. The performance of DUNE is observed to be better than JD, but not around maximal mixing. DUNE and combined setups with DUNE (DUNE + JD and DUNE + JD + KD) exhibit an additional worsening around maximal mixing choices, which is absent in JD and JD + KD. 

To further explore this, we try to understand this nature in individual setups: DUNE and JD. In Fig.~\ref{fig:5}, we represent CP coverage in three scenarios for both JD and DUNE. In the first (dotted blue and red curves), we fix the same value of atmospheric mixing angle in both theory and data; the second (solid blue and red curves) is the result from previously discussed Fig.~\ref{fig:4}, wherein we marginalize in theory over $\sin^{2}\theta_{23}$ in the current 3$\sigma$ uncertain range of [0.4 , 0.6]; in the third (dashed blue and red curves), we show the contribution from only appearance channel marginalized over the allowed $3\sigma$ range in $\sin^2\theta_{23}$. In the fixed parameter case, there is no role of $\theta_{23}-\delta_{\mathrm{CP}}$ degeneracy since both are fixed to their true value in the fit.  Thus, we see monotonically decreasing CP coverage with increasing $\theta_{23}$. This follows the nature of CP asymmetry that we discussed at both probability and event levels. However, once we consider the freedom of uncertainty of $\theta_{23}$ in theory, the CP coverage drastically decreases around the maximal true value of $\tzm$ in the case of DUNE. This signifies that there are CP phases in DUNE, which, when considered in a fixed parameter scenario, gives a 3$\sigma$ or larger sensitivity towards CPV, but they fail to attain the same when the marginalization is performed. So, these act as unfavorable CP phases in DUNE, which are absent in JD. For instance, in DUNE, if we fix both data and theory at $\delta_{\mathrm{CP}}$ = -148$^{\circ}$ and  $\sin^{2}\theta_{23}$ = 0.5, we obtain $\Delta \chi^{2} = 9.82$ (following Eq.~\ref{eq:chi2-cpv}), which drops down to 6.44 when we vary $\sin^{2}\theta_{23}$ in theory; however for the same set of parameters in JD, $\Delta \chi^{2} = 10.6$ changes to $\Delta \chi^{2} = 10.4$ when we change from fixed-parameter to marginalized (over $\sin^{2}\theta_{23}$) scenario. For this reason, in JD, marginalization has a negligible effect on the CP coverage.  Further, there are instances where minimized $\Delta \chi^{2}$ chooses the opposite octant in the fit in the case of DUNE. For instance, when we generate the data with $\delta_{\mathrm{CP}}$ = -145$^{\circ}$ and  $\sin^{2}\theta_{23}$ = 0.48 (LO), we obtain a minimized $\Delta \chi^{2}$ for $\sin^{2}\theta_{23}$ = 0.531 (HO) in DUNE, leading to $\theta_{23}-\delta_{\mathrm{CP}}$ degeneracy, while this is not the case in JD. Therefore this degeneracy becomes crucial when the matter potential is considerable (like in DUNE), while the same degeneracy almost vanishes when we are in a vacuum-like scenario.

Now, the dashed curves represent the contribution from only appearance, which is quite less in DUNE when compared with the CP coverage from both appearance and disappearance (solid red). However, solid and dashed blue-colored curves are almost overlapping. This signifies that the effect of disappearance events is much more crucial in DUNE than in JD. This is because, in JD, the absence of any significant matter effect drastically reduces the impact of the $\theta_{23}-\delta_{\mathrm{CP}}$ degeneracy. Since the leading term in the disappearance channel is dependent on $\sin^{2}2\theta_{23}$, it strongly constrains the $\theta_{23}$ parameter in a $\delta_{\mathrm{CP}}$ independent manner for regions far from maximal mixing. For these values of $\sin^{2}2\theta_{23}$, the appearance channel suffers less from the $\theta_{23}-\delta_{\mathrm{CP}}$ degeneracy in DUNE just like JD. However, for $\sin^{2}\theta_{23}$ around MM, the disappearance rate decreases as the leading term in disappearance is $\propto (1-\sin^{2}\theta_{23})$. Further, as discussed in Fig~\ref{fig:3}, extrinsic or matter-induced fake CP asymmetry also plays an essential role because of the substantial matter effect in DUNE. This worsens the CP coverage since the effect of $\theta_{23}-\delta_{\mathrm{CP}}$ degeneracy is much more dominant. On the other hand, JD remains almost independent of this fake CP asymmetry as the matter effect is negligible here. Therefore, we must notice that despite the bigger systematic uncertainties in T2HK, it can achieve a better CP coverage of true $\dcp$ in leptonic CPV than DUNE around the MM of $\sin^{2}\theta_{23}$. 

For completeness, in Appendix~\ref{sec:appendix1}, Fig.~\ref{fig:5s}, we show the CP coverage for leptonic CP violation at $5\sigma$ as a function of $\sin^2\theta_{23}$. 
%
 \subsection{CP coverage as a function of Exposure}
\label{sec:4b}
%
 \begin{figure}[htb!]
     \centering
     \includegraphics[width=1.0\linewidth]{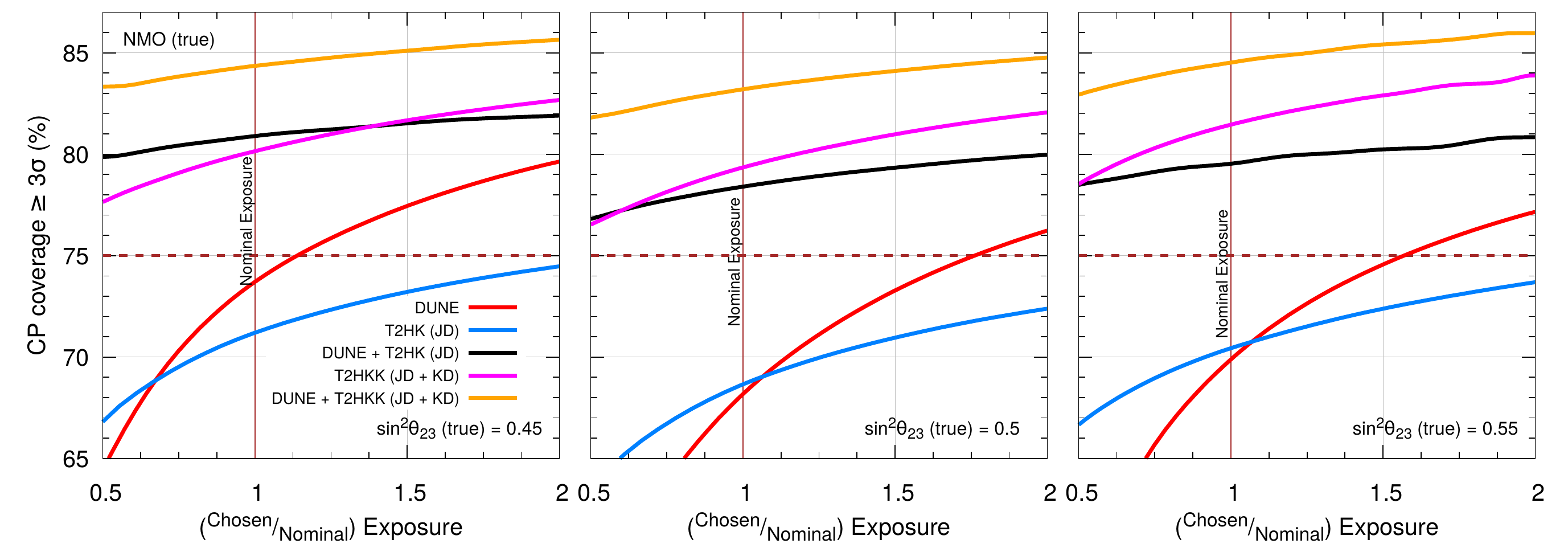}
     \mycaption{Coverage in true $\dcp$ for $\geq 3\sigma$ leptonic CPV as a function of scaled exposure assuming true NMO. We obtain these curves by generating the data with true $\sin^{2}\theta_{23}$ = 0.45 (LO), 0.5 (MM), and 0.55 (HO) and marginalizing over $\theta_{23}$ in the fit in its present 3$\sigma$ range of [0.4 , 0.6], in the left, middle, and right panels, respectively. Here, we define scaled exposure as the ratio of assumed exposure with the nominal exposure of each experiment. Thus, (Chosen/Nominal) Exposure = 1 is the benchmark exposure of the considered setup. The curves: red, blue, black, magenta, and orange correspond to DUNE, JD, DUNE + JD, JD + KD, and DUNE + JD + KD setups, respectively. } 
     \label{fig:6}
 \end{figure}
In this section, we discuss the CP coverage of various experimental setups when the total exposure in the experiment is varied. Recently, the DUNE collaboration had an extensive study on how they expect to achieve desired exposure in a staged manner~\cite{DUNE:2021mtg}. However, to study its effect in our analysis, we take a more simplistic approach and vary the full exposure by reducing it to half of its nominal value and increasing it to twice. In Fig. \ref{fig:6}, we depict the variation in CP coverage on changing the total exposure of the experiment, and fixing $\sin^{2}\theta_{23}=0.45$ (left panel), 0.5 (middle panel), and 0.55 (right panel) in the data. The curves are shown for DUNE (red), JD (blue), DUNE + JD (black), JD + KD (magenta), and DUNE + JD + KD (orange). We marginalize over the atmospheric mixing angle in the fit. We observe that by keeping the true value for $\sin^{2}\theta_{23}$ fixed in LO and doubling the exposure from the nominal value of 2431 (480) kt$\cdot$MW$\cdot$yrs in JD (DUNE), the CP coverage increases from 71\% to 75\% (74\% to 79\% ). On the other hand, if the exposure is reduced to around half the nominal value, the CP coverage drastically reduces for both standalone experiments. However, among both, DUNE outperforms JD when we compare their respective CP coverages at 70\% of their nominal exposures. As expected, we observe that the CP coverage increases with an increase in exposure; however, on comparing LO and MM, we notice that the maximum reachable coverage is now reduced by $\sim (2-3)\%$ in both DUNE and JD. Moreover, under MM, JD outperforms DUNE for nominal exposures. This happens due to strong $\tzm-\dcp$ degeneracy in DUNE near MM, which results in the reduction of CP coverage as compared to JD.
In the HO case, however, the coverage worsens further for JD (maximum coverage 70\%) but not for DUNE. This is because, in HO, DUNE suffers less from the $\theta_{23}-\delta_{\mathrm{CP}}$ degeneracy under the non-maximal case. This follows the previous discussion in Fig.~\ref{fig:6} (dotted red), where we observed that by fixing the value of $\sin^{2}\theta_{23}$, the performance in DUNE improves considerably under MM, but this improvement is very little under the HO case.
 
Interestingly, the complementarity between DUNE and T2HK (JD) achieves more than 75\% CP coverage with just half of their individual nominal exposures in each of the three panels of Fig.~\ref{fig:6}. As discussed earlier, JD + KD further increases the CP coverage significantly for all the three values of true $\sin^{2}\theta_{23}$. Moreover, under the nominal exposure, the combination of DUNE + JD + KD establishes CP coverage always above $\sim 80\%$. 
%
\subsection{Optimizing Runtime for maximal CP coverage}
\label{sec:4c}
%
 \begin{figure}[htb!]
     \centering
     \includegraphics[width=1.0\linewidth]{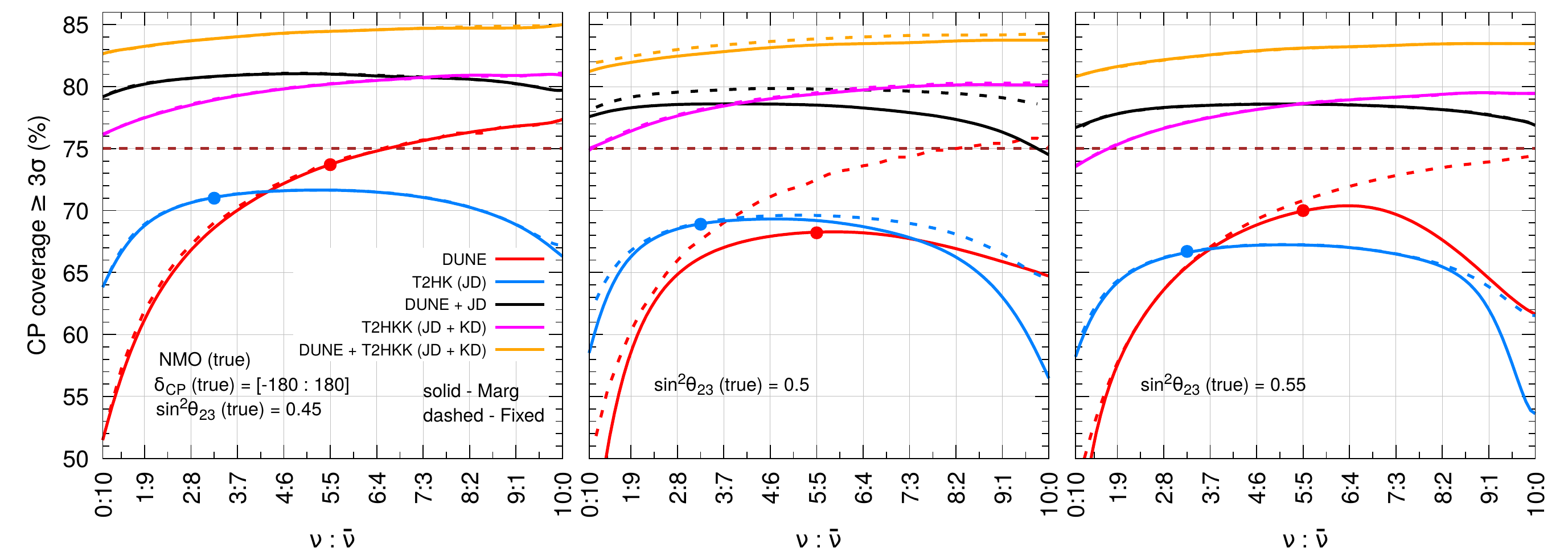}
     \mycaption{Coverage in true $\dcp$ for $\geq 3\sigma$ leptonic CPV as a function of the ratio of the runtime in neutrino and antineutrino ($\nu : \bar{\nu}$) modes. The left, middle, and right panels represent CP coverage with true $\sin^{2}\theta_{23}$ = 0.45 (LO), 0.5 (MM), and 0.55 (HO), respectively. The dashed lines are obtained by considering identical $\sin^{2}\theta_{23}$ in both data and theory, while the solid lines show the result when marginalized over the present 3$\sigma$ uncertain range of $\theta_{23}$ [0.4 , 0.6] in theory. The curves: red, blue, black, magenta, and orange correspond to DUNE, JD, DUNE + JD, JD + KD, and DUNE + JD + KD setups, respectively. The red and blue filled circles depict the nominal runtime in DUNE [5 $\nu$ yrs + 5 $\bar{\nu}$ yrs] and T2HK [2.5 $\nu$ yrs + 7.5 $\bar{\nu}$ yrs], respectively. We assume true NMO in both data and fit.  } 
     \label{fig:7}
 \end{figure}
%
While discussing the total exposure, it is also important to determine the optimal runtime in neutrino and in antineutrino modes for higher CP coverage. The two collaborations have proposed different approaches. 
  With the intent of having a similar number of neutrino and antineutrino events (see in Table~\ref{table:three}), the T2HK (JD) plans to split the total exposure of 10 years into 2.5 years in neutrino and 7.5 years in antineutrino mode (refer to blue filled circles in Fig.~\ref{fig:7}). This choice ensures very small integrated asymmetries in the CP-conserving cases, which help in highlighting easily the differences from the intrinsic asymmetries. Moreover, this choice has already been proven to be very useful for resolving degeneracies~\cite{Agarwalla:2013ju}. Contrastingly in DUNE, they propose a balanced ratio of runtime in neutrino and antineutrino modes of [5 $\nu$ yrs + 5 $\bar{\nu}$ yrs] (refer to red filled circles in Fig.~\ref{fig:7}). Because of the cross-section suppression in antineutrino mode, the number of events in this mode will be considerably lesser than in neutrino mode. However, at the same time, the increment in neutrino events ensures elevated potentials in CP coverage. To visualize this discussion in Fig.~\ref{fig:7}, we represent the optimal ratio between neutrino and antineutrino runtime as we change the value of $\sin^{2}\theta_{23}$ in Nature to establish better CP coverage.  We generate data for three choices of $\sin^{2}\theta_{23}$ (true): LO, MM, and HO. This discussion might shed some light on the choices of runtime in DUNE and JD as we expect the global oscillation data to improve current precision in $\sin^{2}\theta_{23}$ in the coming future.\footnote{In the upcoming years, T2K~\cite{T2K:2023smv} and NO$\nu$A~\cite{Carceller:2023kdz} will complete their data-taking, Super-K~\cite{Posiadala-Zezula:2022vzn}, and IceCube-DeepCore will accumulate more exposure, also, KM3NeT-ORCA~\cite{Lastoria:2023egz} and IceCube Upgrade~\cite{Clark:2021fkg} will further contribute in improving the current precision in $\sin^{2}\theta_{23}$.}. We distinctly show two possible scenarios: first, by fixing the same set of oscillation parameters in both data and theory (dashed curves) and second, by marginalizing $\sin^{2}\theta_{23}$ (solid curves) in theory and fixing all other parameters to the benchmark choices from Table~\ref{table:one}, assuming NMO. In the LO case, while the nominal choice for JD [2.5 $\nu$ yrs + 7.5 $\bar{\nu}$ yrs] turns out to be the best, DUNE has no advantage of running in antineutrino mode. Instead, we observe that the best coverage (77\%) for DUNE is acquired when only neutrino mode is employed for the full 10 years of runtime. This is because of the $\delta_{\mathrm{CP}}$ independent measurement of $\sin^{2}\theta_{23}$ by the disappearance channel in LO. Once the atmospheric angle is constrained by disappearance, the appearance channel benefits more from the increment in statistics by running only in neutrino mode for 10 years instead of a balanced number of neutrino and antineutrino events because of the small appearance systematic uncertainties in DUNE (2\%).
 In the maximal mixing case, the JD remains almost the same, in contrast to DUNE which establishes the best coverage for the balanced runtime scenario. This is because here, the subdued abilities of the disappearance channel in the marginalized $\tzm$ scenario are overcome by the balanced runtime of [5 $\nu$ yrs + 5 $\bar{\nu}$ yrs], thus achieving the best coverage of 68\%. While the HO case in which we generate data by assuming $\sin^{2}\theta_{23}$ (true) = 0.55, we notice that the preferred ratio in runtime is intermediate: the best CP coverage in DUNE is neither obtained by the balanced run, [5 $\nu$ yrs + 5 $\bar{\nu}$ yrs] nor by only surmounting to the highest possible events by running fully in only neutrino mode, instead a runtime of [6.5 $\nu$ yrs + 3.5 $\bar{\nu}$ yrs] scenario is favored. This can be understood from previous discussions around Fig.~\ref{fig:5}. We observe that $\sin^{2}\theta_{23}$ (true) = 0.55 is still in the dip region (refer to solid red in Fig.~\ref{fig:5}) but not completely, thus disappearance is not able to constrain $\sin^{2}\theta_{23}$ in the fit completely, but at the same time does not deteriorate the significance as much as it did in the MM scenario. So, we still feel the effect of the $\theta_{23} - \delta_{\mathrm{CP}}$ degeneracy that requires contribution from both neutrino and antineutrino modes.
 
 However, the complementarity between the two setups plays a crucial role which is essentially independent of $\sin^{2}\theta_{23}$ in Nature. It is quite interesting to observe that DUNE + JD makes the choices of runtime almost irrelevant in establishing coverage in true $\dcp$ for CPV with $\geq$ 3$\sigma$ C.L. around 75\% in all the three panels, given they both run in their full exposure. DUNE + T2HKK can further improve this coverage to about 80\% for the three choices of $\theta_{23}$.
 
 %
  \subsection{Impact of Systematic Uncertainties on CP coverage}
\label{sec:4d}

\begin{figure}[htb!]
     \centering
     \includegraphics[width=1.0\linewidth]{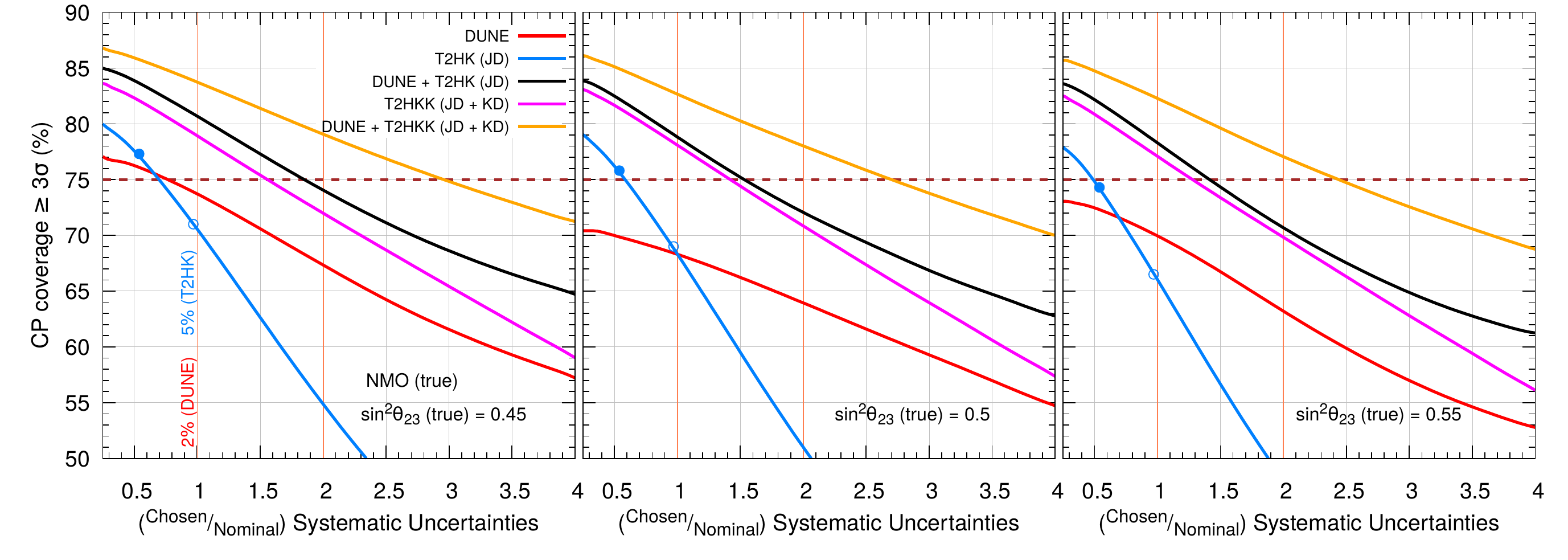}   \mycaption{Coverage in true $\dcp$ for $\geq 3\sigma$ leptonic CPV as a function of scaled appearance systematic uncertainties, assuming NMO. Thus, (Chosen/Nominal) Systematic Uncertainties = 1 refers to the benchmark appearance systematic uncertainties (2\% in DUNE and 5\% in T2HK and T2HKK) of the considered experiment. In the left, middle, and right panels, we obtain the results by considering true $\sin^{2}\theta_{23}$ = 0.45 (LO), 0.5 (MM), and 0.55 (HO), and marginalizing over the 3$\sigma$ uncertain range of $\sin^{2}\theta_{23}$ in the fit, respectively. The curves: red, blue, black, magenta, and orange are for DUNE, JD, DUNE + JD, JD + KD, and DUNE + JD + KD neutrino oscillation experiments, respectively. The blue-colored filled and empty circles in the figure depict CP coverage corresponding to 2.7\% and 4.9\% systematic uncertainties in T2HK, respectively. While the black-filled circles give us the coverage in true $\dcp$ for the combined DUNE + T2HK case, with each setup having 1.5 times its nominal systematic uncertainties. } 
     \label{fig:8}
 \end{figure}
%
In Fig. \ref{fig:8}, we illustrate the effect of appearance systematic uncertainties in the coverage in true $\dcp$ for determining CPV with at least 3$\sigma$ C.L. when marginalized over $\sin^{2}\theta_{23}$ in the fit. It is clear that the JD curve (refer to blue colored curve) has a steeper slope than DUNE (refer to red colored curve); however, one must note that the nominal appearance systematic uncertainties in JD (5\%) is more than twice than that of DUNE (2\%). Recently the T2K collaboration~\cite{T2K:2019bcf} has been considering the conservative uncertainties in the appearance systematics of about 4.9\%, which they further expect to improve to about 2.7\% by the time T2HK starts taking data~\cite{Munteanu:2022zla}. Thus we also discuss these two possibilities in Fig. \ref{fig:8} (refer to blue empty and filled circles). Comparing the CP coverage at the expected T2HK systematics of 2.7\% (filled blue circles) with the nominal in DUNE (2\%), we observe that T2HK outperforms DUNE in the three possible choices of $\sin^{2}\theta_{23}$. We also confirm that in DUNE, the impact of the marginalization becomes negligible  when systematics are higher than 5\%, so the coverage in true $\dcp$ becomes completely systematics-dominated.

Contrastingly, in Nature, if the real appearance systematic uncertainty turns out to be about 1.5 times higher than its nominal value in both DUNE and T2HK setups, then the complementarity between them is the only solution to achieve 75\% of coverage in true $\dcp$ for the three possible choices of $\sin^{2}\theta_{23}$: 0.45, 0.5, and 0.55 (refer to the coordinates of black filled circles in each panel). Also, when we include the second proposed detector: KD, in the analysis along with DUNE (refer to the orange curve), we achieve the milestone of 75\% coverage, even if appearance systematics is increased by a factor of 2.5 for all three possible choices of true $\sin^{2}\theta_{23}$.

\subsection{Effect of Current 3$\sigma$ Allowed Range in $\delta_{\mathrm{CP}}$ on CP coverage}
\label{sec:4e}

 \begin{figure}[t!]
    \centering
    \includegraphics[width = 0.8\linewidth]{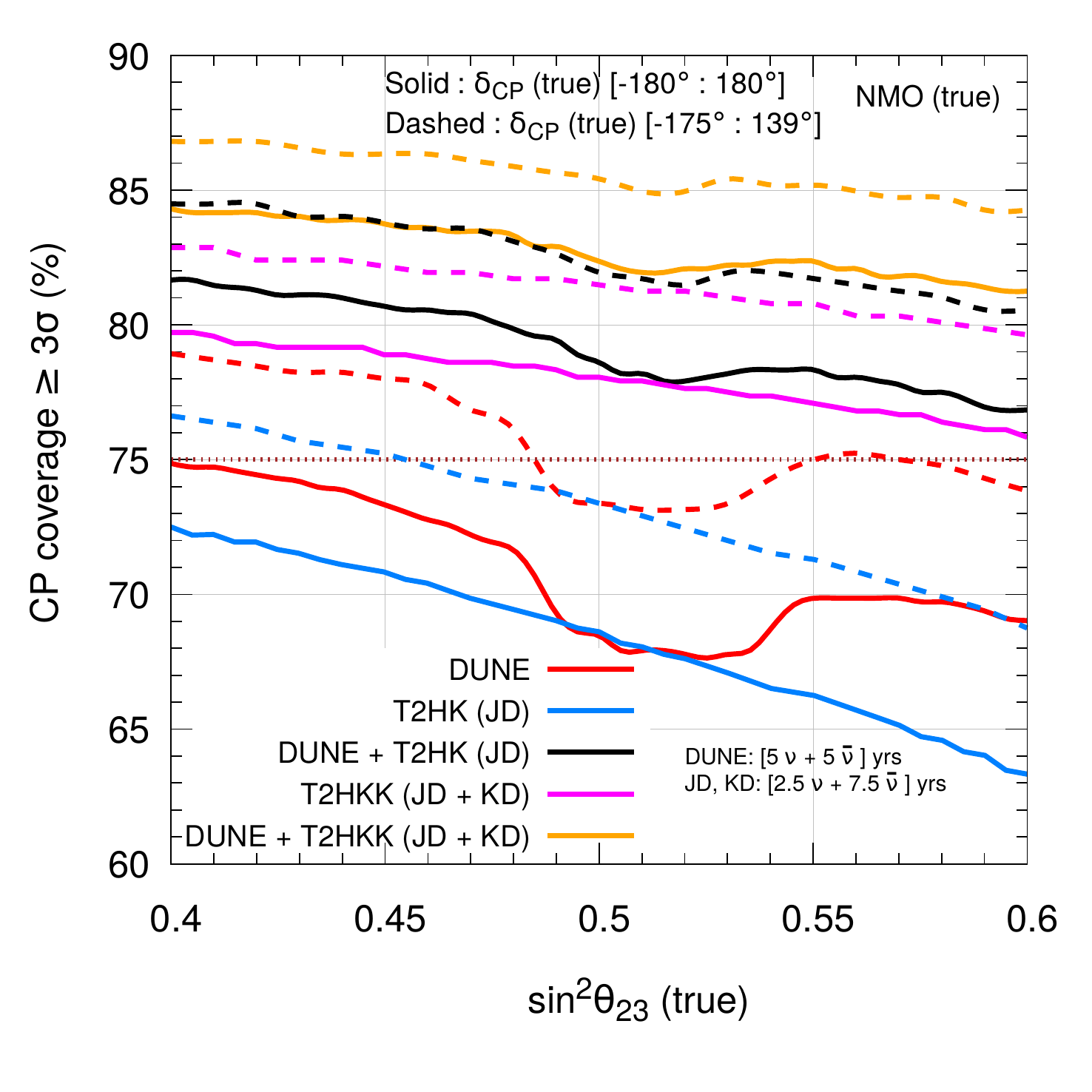}
    \mycaption{Coverage in true $\dcp$ for $\geq$ 3$\sigma$ leptonic CPV as a function of $\sin^{2}\theta_{23}$ (true) while 
    marginalizing over $\theta_{23}$ in the fit. The solid curves represent the results wherein we generate data assuming 
    the entire range of $\delta_{\mathrm{CP}}$ (true) $\in [-180^{\circ} , 180^{\circ}]$ and exclude test $\delta_{\mathrm{CP}}$ 
    = $0^{\circ}$ and $180^{\circ}$ in the fit. On the other hand, the dashed curves show the results when we generate data 
    considering only the present 3$\sigma$ allowed range of $\delta_{\mathrm{CP}}$ (true) 
    $\in [-175^{\circ} , 41^{\circ}]$~\cite{Capozzi:2021fjo} and exclude test $\delta_{\mathrm{CP}}$ = $0^{\circ}$ in the fit. 
    The curves: red, blue, black, magenta, and orange are for DUNE, JD, DUNE + JD, JD + KD, and DUNE + JD + KD setups, 
    respectively. We assume NMO in both data and theory. }
    \label{fig:9}
\end{figure}

In all the previous results, we consider the entire allowed range of $\dcp \in$ [$-180^{\circ} , 180^{\circ}$] for generating 
the prospective data. However, the presently running LBL accelerator experiments: T2K~\cite{T2K:2019bcf} and 
NOvA~\cite{NOvA:2021nfi} along with the high-precision measurement of $\theta_{13}$ from the Daya Bay reactor 
antineutrino experiment~\cite{kam_biu_luk_2022_6683712} have played an important role in constraining the allowed
parameter space of $\dcp$, which is apparent from the global fit analyses of the world neutrino oscillation 
data~\cite{NuFIT,Capozzi:2021fjo,deSalas:2020pgw,Esteban:2020cvm}. Therefore, it becomes imperative to
estimate the CP coverage of the next-generation LBL experiments like DUNE and T2HK, considering the currently 
allowed range of $\dcp$ assuming that the present hints on the allowed values of $\dcp$ will be converted into 
a discovery as more data will become available in the coming years.

In this subsection, we repeat some of our analysis using the present 3$\sigma$ allowed range 
of $\delta_{\mathrm{CP}} \in [-175^{\circ} , 41^{\circ}]$  with a relative 1$\sigma$ uncertainty of 
16\% as obtained in Ref.~\cite{Capozzi:2021fjo}. Since one of the CP-conserving cases, $\dcp=180^\circ$ 
is ruled out in the present $3\sigma$ bound, we now use only $\dcp = 0^\circ$ in the fit and define 
the Poissonian $\Delta \chi^{2}$~\cite{Baker:1983tu} following the frequentist 
approach~\cite{Blennow:2013oma} as follows
\begin{equation}
	\Delta \chi^2 = \underset{(\theta_{23},\,\lambda_1,\, \lambda_2)}{\mathrm{min}} \,\bigg[\chi^{2}(\delta^{\mathrm{true}}_{\mathrm{CP}} \in [-175^{\circ} , 41^{\circ}])-\chi^2(\delta^{\mathrm{test}}_{\mathrm{CP}}= 0^{\circ})\bigg]\,.
	\label{eq:3sig_chi2}
\end{equation}
where $\lambda_1,\,$ and $ \lambda_2$ are the pull parameters in signal and background, respectively. We calculate CP coverage for different experimental setups using Eq.~\ref{eq:3sig_chi2}. In Fig.~\ref{fig:9}, we show coverage in true $\dcp$ which can establish CPV with at least 3$\sigma$ C.L. as a function of $\sin^2\tzm$. The dashed colored curves represent the CP coverage of a given setup calculated using the $3\sigma$ bounds on $\dcp$.
We observe that with better constraints on $\dcp$, we improve the coverage in $\dcp$ consistently for each $\sin^{2}\theta_{23}$ by almost (4 - 5)\% in both DUNE and JD. Previously in the unconstrained scenario (red solid curve), DUNE does not attain the benchmark of 75\% coverage for any value of $\sin^{2}\theta_{23}$; however, with the new definition (dashed red curve) DUNE can attain 75\% of CP coverage for about 58\% of $\sin^{2}\theta_{23}$ in Nature. Therefore, if in Nature $\sin^{2}\theta_{23}$ turns out to be in any value in LO, then DUNE can easily achieve the milestone of 75\% coverage with the nominal appearance systematics and exposure. Similarly, JD, which could achieve only 72\% of CP coverage in the most favorable zone (solid blue curve), improves further to 75\% (dashed blue curve) of coverage if $\sin^{2}\theta_{23} \in [0.4 , 0.45]$ in Nature. 
As expected, the combined setups can achieve enhanced CP coverage. This increment is quite expected, as here we generate data with a more constrained bound on $\dcp$; moreover, in the fit, we consider only one CP-conserving ($\dcp = 0^{\circ}$) value for studying CP violation.

\section{CP coverage assuming Inverted Mass Ordering}
\label{sec:5}
 \begin{figure}[htb!]
     \centering
     \includegraphics[width=0.8\linewidth]{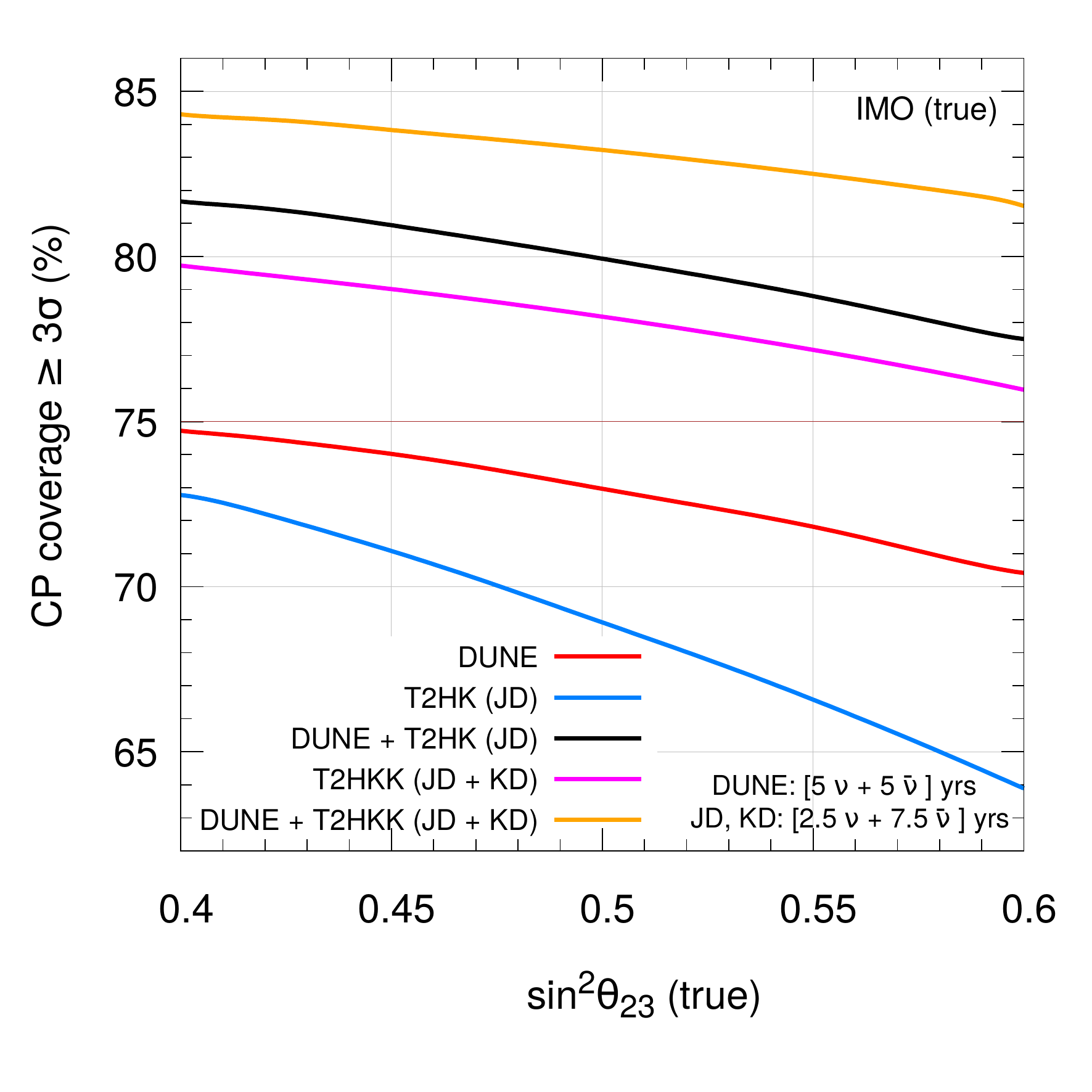}
     \mycaption{Coverage in true $\dcp$ for achieving $\geq 3 \sigma$ leptonic CPV as a function of true $\sin^{2}\theta_{23}$, when marginalized over the current 3$\sigma$ uncertain range of $\sin^{2}\theta_{23}$ [0.4 , 0.6] in the theory. The curves: red, blue, black, magenta, and orange are for DUNE, JD, DUNE + JD, JD + KD, and DUNE + JD + KD neutrino oscillation experiments, respectively. We assume true IMO, benchmark exposure, and the nominal runtime as mentioned in Table~\ref{table:two} in both data and theory.   }
 \label{fig:10}
 \end{figure}

In this section,  we extend the discussion on CP coverage in $\delta_\mathrm{CP}$ for achieving CPV at 3$\sigma$, as a function of true $\sin^{2}\theta_{23}$, assuming IMO. The assumptions and discussion on asymmetries in the appearance channel in Sec.~\ref{sec:2a} at first oscillation maximum (refer to Eq. \ref{FOMmue}) hold true for IMO, with an exception that signs of both $\alpha$ and $\hat A$ become negative. Also, in the case of IMO, the matter effect is felt mostly in the antineutrino mode. Therefore, apart from an overall minus sign and difference in the benchmark value of $\Delta m^{2}_{31}$ (refer to Table~\ref{table:one}), the behavior of the asymmetry in the $\nu_e$ appearance channel, assuming IMO is analogically similar as discussed previously in Sec.~\ref{sec:4a}. Thus, by simply looking at the asymmetries, one would expect the results on the $\dcp$ coverage to be very similar for both mass orderings. Nonetheless, at the probability level, it is possible to notice that in IMO, the $\theta_{23}-\delta_\mathrm{CP}$ degeneracy is relatively milder even in the presence of matter. For instance, we find that if $\Delta m_{31}^2<0$\,, Eq. \ref{rhsb} has no solutions for $\bar{\delta}_{\mathrm{CP}}$ for several values of $\dcp$. This shows that, matter effects in IMO do not allow the data generated in a CP-violating scenario to fit in the hypothesis of the CP-conserving case, for a large range of $\dcp$. Therefore, in IMO, we expect the appearance event rates to be sufficient enough to establish CP violation at a good confidence level even when $\sin^{2}\theta_{23}$ is around MM in DUNE, where matter effects are more pronounced. Further, the increased antineutrino event rates in IMO are also crucial for reducing the statistical uncertainty in the appearance channel of DUNE. The absence of degeneracy is well reflected in Fig.~\ref{fig:10}, where we do not observe any decrease in coverage around $\sin^{2}\theta_{23} = 0.5$ in DUNE. 

Apart from this, the total $3\sigma$ CP coverage for the standalone experiments under IMO is very similar to the NMO scenario, as we expect by studying the nature of the CP asymmetries. For instance, in both Fig.~\ref{fig:10} (assuming true IMO) and Fig.~\ref{fig:4} (assuming true NMO), CP coverage in the leptonic CP violation at 3$\sigma$, decreases roughly from 75\% to 70\% in DUNE. Similarly in T2HK, it reduces from 73\% to 64\% on varying $\sin^2\theta_{23}$ in the range $[0.4,0.6]$. Just like in NMO (refer to Fig.~\ref{fig:4}), in IMO (Fig.~\ref{fig:10}) we find that the combination DUNE + T2HK increases this sensitivity in the range $ 82\% - 77\%$, for $\sin^2\theta_{23} \in [0.4, 0.6]$. While the projected coverage by T2HKK (JD+KD) is comparatively lesser in both NMO and IMO. However, we forecast that the combined DUNE + T2HKK will improve the CP coverage up to 84\% for $\sin^{2}\theta_{23} = 0.4$. For completeness, in Appendix~\ref{sec:appendix1}, in Fig.~\ref{fig:5s}, we compare the coverage for leptonic CPV at $5\sigma$ as a function of $\sin^2\theta_{23}$\,, assuming IMO against NMO.

%
\section{Summary and Conclusions}
\label{sec:6}

The current knowledge of the active three-neutrino mixing angles and two independent 
mass-squared differences has reached unprecedented precision. One of the remaining 
goals is to establish CPV in the leptonic sector. Even though hints of non-vanishing 
$\delta_{\mathrm{CP}}$ are emerging from the current neutrino data, it is worth testing 
the capability of the future long-baseline experiments such as DUNE, T2HK, and T2HKK 
to establish leptonic CPV at $\ge$ 3$\sigma$ C.L. for a large choices of true $\dcp$ 
in its entire range of $[-180^{\circ}, 180^{\circ}]$. Here, we summarize the main findings
of our paper and mention some crucial points that bring out the novelty of this paper.

\begin{itemize}

\item 
In this paper, we extensively discuss the abilities of next-generation high-precision
LBL experiments DUNE, T2HK, and T2HKK in establishing leptonic CPV at $\ge$ 
3$\sigma$ C.L. in isolation and combination, considering their latest state-of-the-art 
configuration details. We emphasize on the fact that DUNE + T2HK is not a mere 
combination of two LBL experiments, but a necessity to achieve the desired 
milestone by reducing their inherent parameter degeneracies that exist in isolation.

\item 
In Sec.~\ref{sec:2a} and Sec.~\ref{sec:2b}, we discuss in detail the intrinsic and 
extrinsic CP asymmetries in appearance and disappearance channels both 
analytically and numerically, explaining their connections to the measurement
of $\delta_{\mathrm{CP}}$. We observe a non-trivial behavior around MM in 
extrinsic CP asymmetry in DUNE, where the matter effect is more important 
and the experiment is less sensitive in measuring $\theta_{23}$ independently 
of the value of $\delta_{\mathrm{CP}}$.

\item 
We observe that assuming NMO, benchmark values of oscillation parameters, 
nominal systematics and exposures, neither DUNE nor T2HK in isolation 
can achieve 75\% CP coverage in true $\dcp$ irrespective of the choices 
of $\theta_{23}$. While the complementarity between DUNE + T2HK can 
enable us to achieve more than 77\% CP coverage irrespective of the 
values of $\theta_{23}$ in its entire 3$\sigma$ allowed range of 
$\sin^2\theta_{23}\in[0.4-0.6]$ (see discussions in Sec.~\ref{sec:4a}).

\item 
We notice that the capabilities of DUNE in establishing leptonic CP violation 
are significantly worse around the maximal mixing value of $\theta_{23}$. 
This is due to the fact that when $\theta_{23}$ is close to $45^{\circ}$, 
the disappearance channel fails to provide a robust measurement of the 
atmospheric mixing angle $\theta_{23}$ independent of the value of 
$\dcp$ (see Fig.~\ref{fig:5}). The underlying physics reason behind 
this is the presence of substantial extrinsic CP asymmetry in the 
disappearance channel around $\sin^{2}\theta_{23}$ $\approx$ 0.5 
in DUNE, which is clearly visible in Fig. 3 and can be easily understood 
from the analytical expression given in Eq.~\ref{eq:4}.

\item
Another interesting observation that we make is that assuming true NMO 
and the benchmark values of oscillation parameters, the CP coverage that 
DUNE + T2HK can achieve with just half of their individual exposures, 
cannot be attained by these experiments in isolation even with twice of 
their nominal exposures for three different choices of true $\theta_{23}$ 
in LO, MM, and HO (see Fig.~\ref{fig:6}). Also, considering our general 
CP coverage estimates as a function of $\theta_{23}$ in Fig.~\ref{fig:4}, 
the results shown in Fig.~\ref{fig:6} as a function of exposure are valid 
for any values of $\theta_{23}$ in its entire 3$\sigma$ allowed range.

\item 
We observe that the different ratios of runtime in neutrino and antineutrino 
modes are needed in DUNE and T2HK to achieve the better CP coverage 
depending upon the values of true $\sin^{2}\theta_{23}$. For instance, 
T2HK always prefers a ratio of $\nu$ and $\bar\nu$ runtimes for which 
the number of appearance events in neutrino and antineutrino modes 
are almost similar irrespective of the choice of true $\sin^{2}\theta_{23}$. 
This is not the case in DUNE. If true $\sin^{2}\theta_{23}$ lies in the 
lower octant, then the expected number of appearance events in DUNE 
becomes very less and therefore, it prefers to have more run in the 
neutrino mode (see discussions in Sec.~\ref{sec:4c}).

\item 
We notice the pronounced effect of larger nominal systematic uncertainties 
in T2HK (5\%) in comparison with DUNE (2\%) in achieving the desired
CP coverage in true $\dcp$ for $\ge$ 3$\sigma$ leptonic CPV. Further, 
we also observe that in a pessimistic scenario in which the systematic 
uncertainties in DUNE and T2HK turn out to be around 1.5 times larger than 
their nominal ones, the combination of the datasets from DUNE and T2HK
is the only solution to achieve the milestone of 75\% CP coverage 
(see discussions in Sec.~\ref{sec:4d}).

\item 
While studying the CP coverage of the T2HKK setup, we discern that the 
combination of the data from both the Japanese and Korean detectors 
(T2HKK/JD+KD) maybe enough to achieve more than 75\% CP coverage 
in true $\dcp$ for $\ge$ 3$\sigma$ leptonic CPV for all the currently allowed 
values of $\theta_{23}$. At the same time, the combination of DUNE and 
T2HKK can attain an unprecedented CP coverage of around 80\% to 85\%
depending upon the value of true $\sin^2\theta_{23}$ (see Fig.~\ref{fig:4}).

\item 
We also study the effect of the currently allowed 3$\sigma$ range in $\dcp$ and 
observe an improvement of about 5\% in CP coverage in both DUNE and T2HK.

\item 
Finally, we also analyze the CP coverage as a function of $\sin^{2}\theta_{23}$ assuming true IMO. We notice that in this case, the $\theta_{23}-\dcp$ degeneracy is milder in the appearance channel. Therefore unlike NMO, we do not observe any decrease in the coverage around maximal mixing in $\sin^{2}\theta_{23}$ using DUNE. Apart from this feature, the projected coverage attainable by the experiments assuming IMO is similar to the NMO case.

\end{itemize}

\begin{table}[h!]
    \centering
    \begin{tabular}{|c|c|c|c|}
    \hline \hline
        \multirow{1.25}{*}{$\sin^{2}\theta_{23}$} & \multicolumn{3}{c|}{Coverage in true $\dcp$ for $\geq$ 3$\sigma$ CPV (\%)}\\
        \cline{2-4}
       (true) & DUNE & T2HK   & DUNE + T2HK \\
        \hline
        0.45 & 73 & 71 (77) & 81  (83)\\
        0.5 & 68 & 69 (76)&  78 (79)\\
        0.55 & 70 & 66 (74) & 79 (80) \\
      \hline \hline
    \end{tabular}
    \mycaption{Coverage in true $\dcp$ for $\ge 3\sigma$ leptonic CPV 
    with nominal exposures and appearance systematic uncertainties 
    in DUNE, T2HK, and DUNE + T2HK for three different true choices 
    of $\sin^{2}\theta_{23}:$ 0.45, 0.5, and 0.55. The data is generated 
    by fixing all the oscillation parameters at their benchmark values 
    as given in Table~\ref{table:one} and varying the true CP phase 
    $\dcp$ in its entire range of [$-$180$^{\circ}$ , 180$^{\circ}$], 
    while in the fit, we marginalize over $\theta_{23}$ in its current 
    3$\sigma$ allowed range assuming NMO both in data and fit.
    The values in the parenthesis have been obtained by considering 
    an improved appearance systematic uncertainty of 2.7\% 
    in T2HK instead of the nominal value of 5\%.}
    \label{table:four}
\end{table}

In Table~\ref{table:four}, we mention the CP coverage in true 
$\delta_{\mathrm{CP}}$ for $\ge$ 3$\sigma$ leptonic CPV 
achievable by DUNE, T2HK, and DUNE + T2HK considering 
their nominal exposures and a systematic uncertainty of 2\% 
(5\%) in $\nu_e/\bar\nu_e$ appearance channel in DUNE (T2HK) in the NMO case.
The bracketed values give the same when an improved systematic 
uncertainty of 2.7\% in $\nu_e/\bar\nu_e$ appearance channel
is considered for T2HK as recently suggested by the 
collaboration~\cite{Munteanu:2022zla}. The complementarity 
between DUNE and T2HK increases the CP coverage achievable 
by DUNE (T2HK) in isolation by about 10\% (13\%) when we 
combine the data from these two experiments. When improved 
systematics are taken into account, we observe a significant 
enhancement in CP coverage by T2HK for the three benchmark 
choices of true $\sin^2\theta_{23}$, outperforming DUNE's performance 
in each case. The combined DUNE + T2HK setup, on the other hand, 
does not exhibit much improvement since the DUNE's contribution 
remains limited by the so-called ($\theta_{23} - \dcp$) degeneracy.
%
\subsection*{Acknowledgments}
%
We thank B. Rebel and J.M. Conrad for useful discussions. S.K.A would like to thank the conveners of the Neutrino Physics Frontier for providing him an opportunity to present the preliminary results from this work in the Snowmass Community Summer Study Workshop at the University of Washington, Seattle, USA during 17th to 26th July, 2022. S.K.A and S.D. acknowledge the support from the Department of Atomic Energy (DAE), Govt. of India, under the Project Identification Number RIO 4001.  S.K.A. is supported by the Young Scientist Research Grant [INSA/SP/YSP/144/2017/1578] from the Indian National Science Academy (INSA). S.K.A. acknowledges the financial support from the Swarnajayanti Fellowship (sanction order No. DST/SJF/PSA- 05/2019-20) provided by the Department of Science and Technology (DST), Govt. of India, and the Research Grant (sanction order No. SB/SJF/2020-21/21) provided by the Science and Engineering Research Board (SERB), Govt. of India, under the Swarnajayanti Fellowship project. S.K.A would like to thank the United States-India Educational Foundation for providing financial support through the Fulbright-Nehru Academic and Professional Excellence Fellowship (Award No. 2710/F-N APE/2021). M.S. would like to thank Swapna Mahapatra for her useful communications. M.S. acknowledges financial support from the DST, Govt. of India (DST/INSPIRE Fellowship/2018/IF180059). The numerical simulations are carried out using the ``SAMKHYA: High-Performance Computing Facility'' at the Institute of Physics, Bhubaneswar, India. 
\appendix
\section{CP coverage for leptonic CP violation at 5$\sigma$}
\label{sec:appendix1} 
\begin{figure}[htb!]
    \centering
    \includegraphics[width=\linewidth]{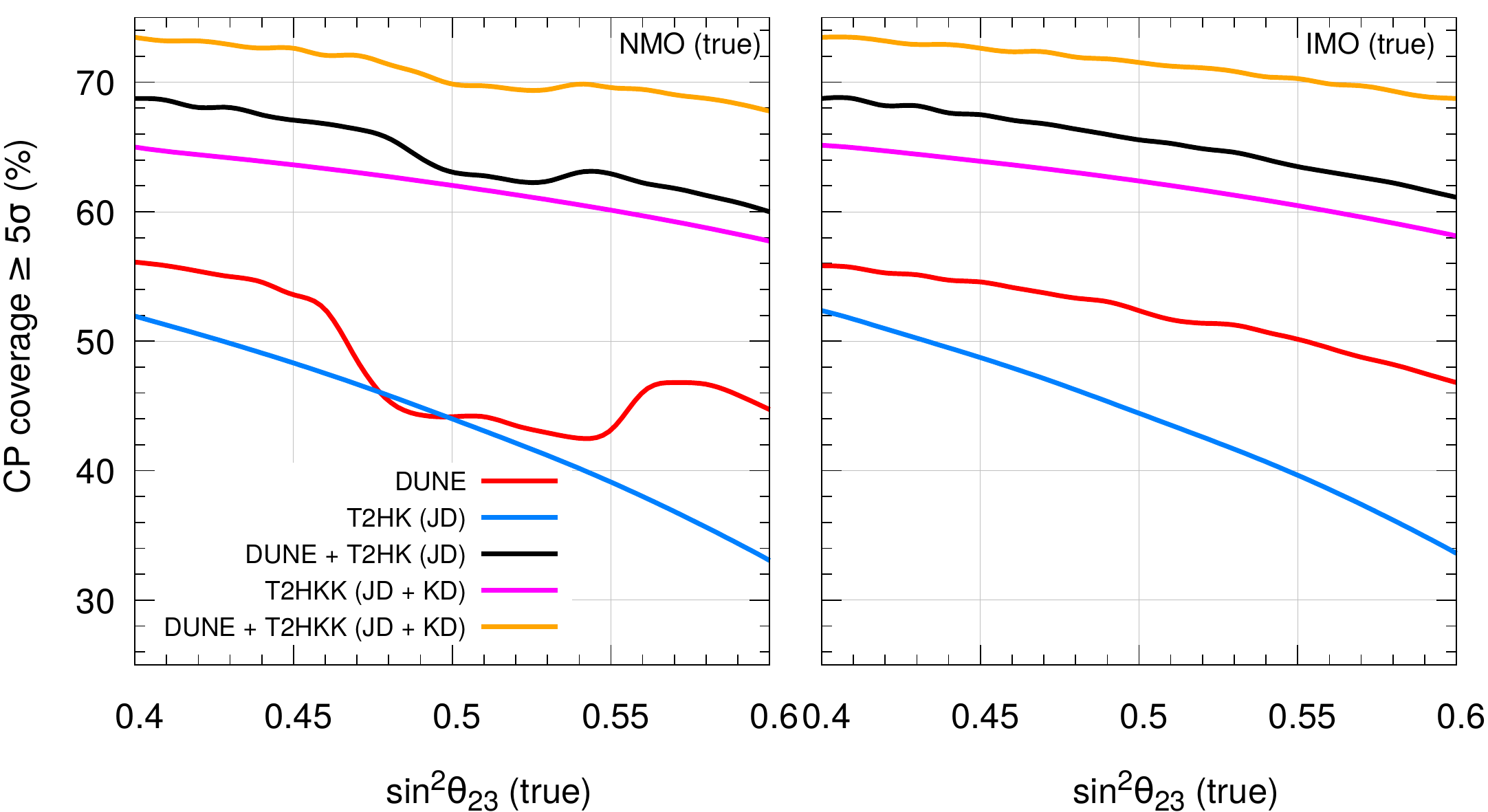}
    \caption{Coverage in true $\dcp$ for achieving $\geq 5 \sigma$ leptonic CPV as a function of true $\sin^{2}\theta_{23}$, assuming true NMO (IMO) in the left (right) panel, when marginalized over the current 3$\sigma$ uncertain range of $\sin^{2}\theta_{23}$ [0.4 , 0.6] in the theory. The curves: red, blue, black, magenta, and orange are for DUNE, JD, DUNE + JD, JD + KD, and DUNE + JD + KD neutrino oscillation experiments, respectively. We assume benchmark exposure and the nominal runtime as mentioned in Table~\ref{table:two} in both data and theory.}
    \label{fig:5s}
\end{figure}
The nature of curves in Fig.~\ref{fig:5s} follows the tendency we observed in Fig.~\ref{fig:4} and Fig.~\ref{fig:10} assuming NMO and IMO, respectively. In Fig.~\ref{fig:5s}, the maximum CP coverage attained by DUNE (T2HK) is 56\% (52\%) and the minimum is 45\% (33\%). On the other hand, a projected 5$\sigma$ discovery of CP violation is achievable for $\sim 60\%$ CP phase, irrespective of the mass ordering and $\sin^{2}\theta_{23}$ in Nature, on considering the combined DUNE + T2HK setup. Further, the projected discovery potential of DUNE + T2HK is better than T2HKK, hence reducing the impact of the second detector in T2HK. However, DUNE + T2HKK could raise the 5$\sigma$ coverage up to 74\%, for $\sin^{2}\theta_{23} = 0.4$.

\bibliographystyle{JHEP}
\bibliography{refrences-cp-coverage}

\end{document}